\documentclass[aps,prd,10pt,showpacs,amsmath,showkeys,twocolumn,floatfix,amssymb, preprintnumbers, nofootinbib, superscriptaddress]{revtex4}
\usepackage{graphicx}
\usepackage{comment}
\usepackage[usenames]{color}
\usepackage{bm}
\usepackage{ifpdf}
\usepackage{floatrow}
\usepackage{makecell}
\usepackage{caption}

\usepackage[normalem]{ulem}
\usepackage[dvipsnames]{xcolor}
\usepackage[utf8]{inputenc}
\usepackage{hyperref}
\hypersetup{
  pdfnewwindow=true,      
  colorlinks=true,        
  linkcolor=PineGreen,    
  citecolor=PineGreen,    
  filecolor=PineGreen,    
  urlcolor=PineGreen      
}
\newcommand{\be}{\begin{eqnarray}}
\newcommand{\ee}{\end{eqnarray}}

\begin{document}

\title{Friedel formula and Krein's theorem in complex potential scattering theory}

\author{Peng~Guo}
\email{pguo@csub.edu}

\affiliation{Department of Physics and Engineering,  California State University, Bakersfield, CA 93311, USA}

\author{Vladimir~Gasparian}
\email{vgasparyan@csub.edu}

\affiliation{Department of Physics and Engineering,  California State University, Bakersfield, CA 93311, USA}

\date{\today}

\begin{abstract}
In this work, the generalization of Friedel formula and Krein's theorem in  complex potential scattering theory  is presented. The consequence of various   symmetry constraints on dynamical system are   discussed. In addition, Muskhelishvili-Omn\`es representation of Krein's theorem is also given and discussed.
\end{abstract}

\maketitle

\section{Introduction}\label{sec:intro}

A remarkable relation that connects the     integrated density of states  of a system  and  energy derivative of scattering phaseshifts was given by J. Friedel in  Refs.~\cite{doi:10.1080/00018735400101233,Friedel1958}, which   is   referred as the Friedel formula and find wide applications in solid states, multiple-scattering theory, etc.  The  similar   relation  was also derived and  found uses in statistical mechanics \cite{PhysRev.187.345,OSBORN1976119}.   The Friedel formula  was originally used to describe the change of density of states due to the perturbation of impurity placed in the metal. Integrating both sides of the Friedel formula   over the  energy up to the Fermi energy, it leads to the well-known Friedel sum rule \cite{Friedel1958}, which relates the total charge of screening  conduction electrons around a charged impurity to the scattering phaseshifts. The generalization of the Friedel formula  into  multiple-scattering theory in the calculation of electronic band structure results in another well-known relation:  Lloyd formula  \cite{Lloyd_1967}.   Other important  modern applications of    Friedel formula    include the development of the concept of time-delay in collision theory, see Refs.\cite{PhysRev.98.145,PhysRev.118.349,goldberger2004collision}, where the integrated  density of states  is   interpreted as the lifetime of scattering states tunneling through potential barriers and usually  referred as Wigner time-delay.  Later on, it was recognized by J.S. Faulkner in Ref.~\cite{Faulkner_1977} that Friedel formula can be derived from Krein's theorem \cite{zbMATH03313022} in spectral theory.

The aim of present work is to explore the possibility of generalization of Friedel formula and Krein's theorem when interaction potential is complex, and also study what other new features the complex potential scattering theory may bring in? All the discussions are currently confined only in one dimensional space. We will show later on that even with complex potentials in general,  the similar relations as Friedel formula and Krein's theorem in real potential theory can be obtained. Though the physical interpretation of such relations in complex potentials scattering theory may be drastically different from those in real potential scattering theory. In real potential scattering theory, the conservation of norm of states plays the central role in interpreting the absorptive part of Green's function as density of states of system. On the contrary, in complex potential scattering theory, because of absorbing or emissive nature of complex potential, the norm of states is no longer conserved. However,  for    dual  systems with two subsystems, one is absorbing with loss and another is emissive with gain, when the gain and loss of   dual systems are balanced in dynamic equilibrium,  the biorthogonal relation between the eigenstates of dual systems can be established.  Due to the resemblance of biorthogonal basis in non-Hermitian theory and orthogonal basis  in Hermitian theory,  Friedel formula and Krein's theorem type of relations in complex potential scattering theory maintain the similar mathematical forms, but  the absorptive part of Green's function is no longer related to the density of states,  and it is a complex function in general.  Similar mathematical forms of  Friedel formula and Krein's theorem type of relations in complex potential scattering theory   are the consequence of the balanced gain and loss in dual systems. Only in the special cases, such as $\mathcal{P}\mathcal{T}$ symmetric systems, the absorptive part of Green's function may still be  real, though positive-definite of norm is not guaranteed. Therefore, in collision theory, the absorptive part of $\mathcal{P}\mathcal{T}$ symmetric Green's function may   be  interpreted as the generalized time delay of particle scattering off $\mathcal{P}\mathcal{T}$ symmetric complex barriers. The positivity and negativity of generalized time delay simply reflect the nature of potential barriers that either tend to  keep particle in or force it out. 

Such a study is primarily motivated by recent advance in both experimental and theoretical developments in study of $\mathcal{P}\mathcal{T}$ symmetric systems, see Refs.~\cite{doi:10.1142/q0178,RevModPhys.88.035002,doi:10.1080/00107500072632,doi:10.1142/S0219887810004816,Mostafazadeh2009,El-Ganainy2018,PhysRevLett.100.103904,PhysRevLett.100.030402}. Especially the experimental realization of $\mathcal{P}\mathcal{T}$ symmetric systems in optics \cite{El-Ganainy2018,PhysRevLett.100.103904,PhysRevLett.100.030402},  atomic gases \cite{PhysRevLett.110.083604,Hang:14}, plasmonic waveguides \cite{PhysRevA.89.033829,PhysRevB.89.075136}, acoustic \cite{PhysRevX.4.031042}, etc. may make it feasible for the study of some interesting  subjects, such as tunneling time of a particle through complex barriers, multiple scattering theory in $\mathcal{P}\mathcal{T}$ symmetric systems, etc.       Many intriguing processes take place in  photonic systems with unbroken $\mathcal{P}\mathcal{T}$ symmetry or in $\mathcal{P}\mathcal{T}$ symmetry breaking phase. For instance, in $\mathcal{P}\mathcal{T}$ symmetric crystals, the violation of the normal conservation of the photon flux   leads to anisotropic transmission resonances \cite{PhysRevA.85.023802}. 
In $\mathcal{P}\mathcal{T}$ symmetry breaking phase, the optical reciprocity yields the unity of the product of the two eigenvalues of the scattering matrix   \cite{PhysRevA.94.043836}, consequently double refraction and unidirectional invisibility become possible. This may have the significant impact on the dwell time of particle tunneling through barriers, which is conventionally defined as a weighted average between both   transmission  and reflection   times.

The paper is organized as follows.  A brief summary of    Friedel formula and Krein's theorem in real potential scattering theory is provided in Sec.~\ref{summFriedelformula}. The derivation of Friedel formula and Krein's theorem in complex potential scattering theory are presented in Sec.~\ref{Fridelformula} and Sec.~\ref{Kreintheorem} respectively, followed by   the discussions and summary   in Sec.~\ref{summary}.  A short introduction  of   complex potential scattering theory and    discussion of symmetry constraints  are   also provided in Appendix~\ref{scattgen} and Appendix~\ref{scattPTpot} respectively  for readers who may not be familiar with complex potential scattering theory.

\section{Summary of Friedel formula and Krein's theorem in a real potential scattering theory}\label{summFriedelformula}
In the real potential scattering theory, the local density of states of a system, $n_E(x)$, is related to the imaginary part of Green's function by
\begin{equation}
n_E (x) = - \frac{1}{\pi} Im  \left [ \langle x | \hat{G}(E+ i 0 )| x \rangle \right ],
\end{equation}
where $$\hat{G}(E) = \frac{1}{E- \hat{H}}$$ refers to full  Green's function operator of an interacting system, and   $\hat{H}  $ stands for the Hamiltonian operator of the  system.  
 The spectral representation of Green's function  has the form of
\begin{equation}
\hat{G} (E) =  \sum_\epsilon \frac{ | \Psi_\epsilon  \rangle \langle  \Psi_\epsilon   |}{E- \epsilon}, \label{Gspectralsum}
\end{equation}
where $| \Psi_\epsilon  \rangle$ are   eigenstates of Hamiltonian $\hat{H}$: 
$\hat{H} | \Psi_\epsilon  \rangle = \epsilon | \Psi_\epsilon  \rangle $, and spectrum sum in Eq.(\ref{Gspectralsum}) includes the  sum of both discrete bound states and continuous scattering states. The normalization and completeness of eigenstates,
$$ \sum_\epsilon  | \Psi_\epsilon  \rangle \langle  \Psi_\epsilon   | = \mathbb{I},$$
 warrants  the interpretation of  imaginary part of Green's function as the density of state
\begin{equation}
n_E (x) =     |  \langle x |  \Psi_{E+ i 0} \rangle |^2.
\end{equation}

In Refs.~\cite{doi:10.1080/00018735400101233,Friedel1958}, J. Friedel showed that the difference between the integrated density of states of the interacting system and free particle system, $n_E^{(0)} (x)$, is related to the scattering phaseshifts by
\begin{equation}
\int_{-\infty}^\infty d x \left [n_E (x) - n_E^{(0)} (x) \right ] = \frac{1}{ \pi} \frac{d }{d E} Tr \left [ \delta (k) \right ], \label{Friedelformula}
\end{equation}
where $\delta(k)$ stands for the diagonal matrix of scattering phaseshifts, and $k$ is related to mass and energy of scattering particle, $m$ and $E$ respectively, by $$k^2= 2 m E.$$ We remark that  both $k$ and $E$ are used to label the energy dependence of a physical quantity throughout the entire presentation,   the purpose is solely for the convenience of presentation. The relation given in 
Eq.(\ref{Friedelformula}) sometime is referred as the Friedel formula, and  it is usually also given in terms of $S$-matrix, $$S(k) =e^{2 i \delta(k)},$$  by
\begin{align}
&  - \frac{i}{2 \pi } \frac{d}{d E}\ln \det \left [  S(k) \right ] \nonumber \\
&=   Im  \left [  \int_{-\infty}^\infty d x  \langle x |  \hat{G}(E+ i 0 )  -  \hat{G}_0(E+ i 0 ) | x \rangle\right ]  ,\label{SintegratedG}
\end{align}
where $$\hat{G}_0(E) = \frac{1}{E- \hat{H}_0}$$ denotes the    free particle's Green's function operator.

Given the fact that  Green's functions   has a physical branch cut along the positive real axis in complex $E$-plane: $E \in [0, \infty]$, the physical observables, such as  density of states, phaseshifts and $S$-matrix, etc., are all defined right above the physical branch cut. In addition, the Green's function may also have an unphysical branch cut siting along negative real axis: $E \in [- \infty, -E_L]$,  where $-E_L$ represents the branch point of unphysical cut.   In the unphysical region, though  Eq.(\ref{SintegratedG}) is still formally valid, the $S$-matrix and scattering amplitudes  are usually not well constrained and largely model dependent.   The imaginary part (absorptive part) of Green's function is identical to the discontinuity of Green's function across the   physical and unphysical branch cuts, which is given in Eq.(\ref{SintegratedG}). The real part (principal part) of Green's function   can be constructed through imaginary part by Cauchy's integral theorem (also referred as dispersion integral relation in nuclear/particle physics), hence,
\begin{align}
&   \int_{-\infty}^\infty d x   \langle x | \hat{G}(E )  -  \hat{G}_0(E )| x \rangle    \nonumber \\
& =  \frac{d}{d E} \frac{i}{2\pi} \left [ \int_{-\infty}^{-E_L} +\int_{0}^\infty \right ] d \epsilon \frac{\ln \det \left [ S( \sqrt{2 m \epsilon}) \right ]}{\epsilon - E} . \label{Kreinformula}
\end{align}
The equivalence of relation given in Eq.(\ref{Kreinformula}) and Krein's theorem \cite{krein1953trace,zbMATH03313022} in spectral theory is recognized by J.S. Faulkner \cite{Faulkner_1977}, where 
$ \frac{i}{2 \pi } \ln \det \left [  S(k) \right ] $ is exactly the  Krein's spectral shift function, see Ref.~\cite{zbMATH03313022}.  In the collision theory, $ - i \frac{d}{d E} \ln \det \left [  S(k) \right ] $ is also used to describe the time delay of a scattered particle off potential barriers.

In Sec.~\ref{Fridelformula} and Sec.~\ref{Kreintheorem}, we will show that even in complex potential scattering,  the Friedel formula and Krein's theorem remain the similar forms as those in Eq.(\ref{SintegratedG}) and Eq.(\ref{Kreinformula})  in real potential scattering theory. However, the imaginary part of Green's function must be replaced by the absorptive part of Green's function. In complex potential scattering theory, the spectral representation of Green's function  is given in terms of biorthogonal basis of dual systems with balanced gain and loss: one is absorbing with loss and another is emissive with gain.  Hence the absorptive part of Green's function in complex potential scattering   is no longer related to the  density of states of a single system.

\section{Friedel formula in complex potential scattering theory}\label{Fridelformula} 
In this section,  we   show in great details      the   derivation of generalizing  Friedel formula in Eq.(\ref{SintegratedG})   in the complex potential scattering theory. The derivation can be made in a rather more general but intuitive way following the discussion and approach presented in Ref.~\cite{PhysRev.187.345}.

 First of all, considering that a non-relativistic spinless particle of mass $m$ is scattered off a  complex absorbing potential, the dynamics is thus described by Schr\"odinger equation,
 \begin{equation}
 \hat{H} | \Psi_E \rangle  = E  | \Psi_E   \rangle  , \ \ \ \  \hat{H}=\hat{H}_0 + \hat{V}, \label{schrodingereq}
 \end{equation}
 where $\hat{H}_0 =  - \frac{1}{2 m} \frac{d^2}{d x^2}$  and $\hat{V}$  stand for free Hamiltonian  and complex absorbing potential operators of the   system respectively. Its dual system  with   an adjoint Hamiltonian $\hat{H}^\dag$ thus describes a particle scatters off  an emissive complex potential of $\hat{V}^\dag$,  thus the dynamics of the emissive system is given by Schr\"odinger equation,
  \begin{equation}
  \hat{H}^\dag  | \widetilde{ \Psi}_E   \rangle   = E |   \widetilde{ \Psi}_E   \rangle .
 \end{equation}
 The emissive system with gain can be considered as   the time-reversed version of the absorbing system with equal but opposite loss, and vice versa. Hence dual systems have no net gain or loss.
The wavefunction of an absorbing system and its dual are defined in Hilbert space $\mathcal{H}$ and its dual space $\mathcal{H}^*$ respectively, and they are related by
  \begin{equation}
 | \Psi_E \rangle  \leftrightarrow     \langle \widetilde{ \Psi}_{E^*}  |. \label{wavdualrel}
 \end{equation}
 The eigenstates of neither an absorbing nor an emissive system alone form a   orthogonal basis, however the eigenstates of  dual systems together are biorthogonal and normalized as, see Refs.~\cite{FESHBACH1985398,MUGA2004357,Brody_2013},  
  \begin{equation}
\sum_E | \Psi_E   \rangle   \langle \widetilde{ \Psi}_{E}  | =  \mathbb{I} . \label{wavdualnorm}
 \end{equation}
 The expectation value of a observable  $\hat{\mathcal{O}}$  is defined by
 \begin{equation}
  \langle   \hat{\mathcal{O}}    \rangle=   \langle \widetilde{ \Psi}_{E}  |  \hat{\mathcal{O}}  | \Psi_E   \rangle .
 \end{equation}

\subsection{$S$-matrix and Moller operators in complex potential scattering theory}  
 The $S$-matrix operators for dual systems are defined by, see Refs.~\cite{FESHBACH1985398,MUGA2004357},
 \begin{equation}
 \hat{S}(E) = \hat{\widetilde{\Omega}}^{\dag}_{E - i 0} \hat{\Omega}_{E + i 0} ,  \ \ \ \   \hat{\widetilde{S}}(E) = \hat{\Omega}^{\dag}_{E - i 0}  \hat{ \widetilde{\Omega}}_{E + i 0}, \label{SMollerdef}
 \end{equation}
 where Moller operators are defined through wavefunctions by
 \begin{equation}
 | \Psi_E \rangle =\hat{ \Omega}_{E }  | \Psi^{(0)}_E \rangle
 \end{equation}
 for an absorbing system with loss, and
  \begin{equation}
 \langle \widetilde{ \Psi}_E  | =   \langle \Psi^{(0)}_E  |  \hat{\widetilde{\Omega}}^\dag_{E }
 \end{equation}
 for an emissive system with gain respectively. $| \Psi^{(0)}_E \rangle$ stands for the eigenstate of free Hamiltonian, 
 \begin{equation}
 \hat{H}_0 | \Psi^{(0)}_E \rangle = E | \Psi^{(0)}_E \rangle.
 \end{equation}
Moller operators, $\hat{ \Omega}_{E+i0 }$ and $\hat{\widetilde{\Omega}}_{E-i0 }$,  hence describe systems evolve   forward in time with $\hat{H}$ and   backward in time with $\hat{H}^\dag$ respectively.
As the consequence of balanced gain and loss in dual systems,  biorthogonal eigenstates of dual systems are normalized according to
\begin{equation}
 \langle \widetilde{ \Psi}_E | \Psi_E \rangle  = \mathbb{I}
\end{equation}
hence it yields
\begin{equation}
 \hat{\widetilde{\Omega}}^\dag_{E } \hat{\Omega}_{E }     = \mathbb{I}. \label{normMollerops}
\end{equation}
The unitarity relation of $S$-matrix operator is also warranted:
\begin{equation}
\hat{\widetilde{S}}^\dag (E)     \hat{S}(E)   = \hat{ \widetilde{\Omega}}^{\dag}_{E + i 0}  \hat{ \Omega}_{E - i 0} \hat{\widetilde{\Omega}}^{\dag}_{E - i 0} \hat{\Omega}_{E + i 0}  = \mathbb{I}.
\end{equation}
Using Eq.(\ref{SMollerdef}), we also find
\begin{align}
& Tr \left [ \hat{\widetilde{S}}^\dag (E)   \frac{d}{d E}   \hat{S}(E)  -  \hat{S}(E)  \frac{d}{d E}  \hat{\widetilde{S}}^\dag (E)   \right ] \nonumber \\
& = Tr \left [     \hat{ \Omega}_{E - i 0}   \frac{d}{d E} \hat{\widetilde{\Omega}}^{\dag}_{E - i 0}  - \hat{\widetilde{\Omega}}^{\dag}_{E - i 0}   \frac{d}{d E}  \hat{ \Omega}_{E - i 0}    \right ] \nonumber \\
& + Tr \left [    \hat{ \widetilde{\Omega}}^{\dag}_{E + i 0}     \frac{d}{d E} \hat{\Omega}_{E + i 0} -  \hat{\Omega}_{E + i 0}  \frac{d}{d E}  \hat{ \widetilde{\Omega}}^{\dag}_{E + i 0}     \right ] . \label{SOmegaeq}
\end{align}

Next, before we start  simplifying Eq.(\ref{SOmegaeq}),  let's make a list of some useful equations for complex scattering systems.  The Lippmann-Schwinger  (LS) equation for an absorbing system
\begin{equation}
 | \Psi_E \rangle  = | \Psi^{(0)}_E \rangle + \hat{G}_0 (E)\hat{V} | \Psi_E \rangle 
\end{equation}
yields
\begin{equation}
 \hat{ \Omega}_{E }  = \mathbb{I}   - \hat{G}_0 (E) \hat{T} (E) ,   \label{OmegaT}
\end{equation}
where $\hat{T} (E)$ stands for scattering amplitude operator and is defined by
\begin{equation}
  \hat{T} (E) = - \hat{V}  \hat{ \Omega}_{E } .   \label{TVOmega}
\end{equation}
Also using the relation between wavefunction and full Green's function of an absorbing system,
\begin{equation}
 | \Psi_E \rangle  = | \Psi^{(0)}_E \rangle +  \hat{G} (E) \hat{V} | \Psi^{(0)}_E \rangle  , 
\end{equation}
the Moller operator $ \hat{ \Omega}_{E } $ is hence also given by
\begin{equation}
 \hat{ \Omega}_{E }  = \mathbb{I} +   \hat{G} (E) \hat{V},   \label{OmegaGV}
\end{equation}
The scattering amplitude operator, $\hat{T} (E)$, is related to Green's function by
\begin{equation}
 \hat{T} (E)   = - \hat{V} -   \hat{V} \hat{G} (E) \hat{V} .     \label{TGrelation}
\end{equation}
The normalization of Moller operators in dual systems in Eq.(\ref{normMollerops}) and Dyson equation,
\begin{equation}
\hat{G} (E) = \hat{G}_0 (E) + \hat{G}_0 (E) \hat{V}\hat{G} (E) , \label{Dysoneq}
\end{equation}
suggest that
\begin{equation}
 \hat{\widetilde{\Omega}}^\dag_{E }    =   \hat{\Omega}^{-1}_{E }  =  \mathbb{I} -   \hat{G}_0 (E) \hat{V}. \label{OmegaInverse}
\end{equation}

Now  we are ready to  simplify Eq.(\ref{SOmegaeq}) and derive Friedel  formula for complex potential scattering systems.  Using       Eq.(\ref{TVOmega}) and Eq.(\ref{OmegaInverse}),  we first find
\begin{align}
&  Tr \left [     \hat{ \Omega}_{E  }   \frac{d}{d E} \hat{\widetilde{\Omega}}^{\dag}_{E  }  - \hat{\widetilde{\Omega}}^{\dag}_{E  }   \frac{d}{d E}  \hat{ \Omega}_{E  }    \right ] \nonumber \\
&  = Tr \left [ -  \hat{ \Omega}_{E  }  \frac{d}{d E} \hat{G}_0 (E)  \hat{V} - \frac{d}{d E}  \hat{ \Omega}_{E  }   + \hat{G}_0 (E) \hat{V} \frac{d}{d E}  \hat{ \Omega}_{E  }    \right ]  \nonumber \\
&  = Tr \left [  \hat{ T} (E) \frac{d}{d E} \hat{G}_0 (E)  - \frac{d}{d E}  \hat{ \Omega}_{E  }   - \hat{G}_0 (E)   \frac{d}{d E}  \hat{ T}(E  )    \right ]  . 
\end{align}
Next using Eq.(\ref{OmegaT}), we can write  it again to
\begin{align}
&  Tr \left [     \hat{ \Omega}_{E  }   \frac{d}{d E} \hat{\widetilde{\Omega}}^{\dag}_{E  }  - \hat{\widetilde{\Omega}}^{\dag}_{E  }   \frac{d}{d E}  \hat{ \Omega}_{E  }    \right ] \nonumber \\
& =2  Tr \left [  \hat{ T} (E) \frac{d}{d E} \hat{G}_0 (E)     \right ]    =- 2  Tr \left [ \hat{G}_0 (E)   \hat{ T} (E)  \hat{G}_0 (E)     \right ] .\label{OmegaG0TG0}
\end{align}
Finally    Eq.(\ref{TGrelation}) and Dyson equation yields
\begin{equation}
 - \hat{G}_0 (E)  \hat{T} (E)   \hat{G}_0 (E)=  \hat{G} (E) \hat{V} \hat{G}_0 (E) , 
\end{equation}  
 hence     Eq.(\ref{OmegaG0TG0}) can be rewritten further to
  \begin{equation}
  Tr \left [     \hat{ \Omega}_{E  }   \frac{d}{d E} \hat{\widetilde{\Omega}}^{\dag}_{E  }  - \hat{\widetilde{\Omega}}^{\dag}_{E  }   \frac{d}{d E}  \hat{ \Omega}_{E  }    \right ]     = 2  Tr \left [ \hat{G} (E) \hat{V}  \hat{G}_0 (E)     \right ] . \label{OmegaGVG0}
\end{equation}
In the end, Eq.(\ref{SOmegaeq}) and  Eq.(\ref{OmegaGVG0}) together  lead to
 \begin{align}
&- \frac{1}{2}  Tr \left [\hat{ \widetilde{S}}^\dag (E)   \frac{d}{d E}   \hat{S}(E)  -  \hat{S}(E)  \frac{d}{d E}   \hat{\widetilde{S}}^\dag (E)   \right ] \nonumber \\
& =    Tr \left [       \hat{G} (E+i0) \hat{V} \hat{G}_0 (E+i0)    -  \hat{G} (E-i0) \hat{V} \hat{G}_0 (E-i0)      \right ]    . \label{FriedelSGVG0}
\end{align}
The Eq.(\ref{FriedelSGVG0}) holds in general for an arbitrary complex potential without any symmetry consideration. However, in general
$$  \hat{G} (E) \hat{V} \hat{G}_0 (E)  \neq \hat{G}_0 (E) \hat{V}  \hat{G} (E)  = \hat{G} (E)  -\hat{G}_0 (E), $$ the Dyson equation in Eq.(\ref{Dysoneq}) for an arbitrary complex potential is in fact direction dependent. The transpose of Dyson equation
$$    \hat{G} (E) \hat{V} \hat{G}_0 (E) =  \hat{G} (E) -\hat{G}_0 (E)   $$ is valid only if $$ \hat{G}^T (E) = \hat{G} (E) .$$ This is indeed the case when reciprocity symmetry is satisfied under the condition: $\hat{V}^T = \hat{V}$, see Refs.\cite{PhysRevA.78.064101,doi:10.1063/1.1704136,DILLON1968623}.

\subsection{Friedel  formula under symmetry constraints} 
For the local complex potentials, the reciprocity symmetry is automatically satisfied: $$\hat{V}^T = \hat{V}.$$ It can be easily show   \cite{PhysRevA.78.064101,doi:10.1063/1.1704136,DILLON1968623}  that  the Green's function is reciprocal symmetric under the exchange of variables,
\begin{equation}
\langle x | \hat{G} (E) | x' \rangle = \langle x' | \hat{G} (E) | x \rangle.
\end{equation}
Therefore, the Dyson equation is now also reciprocal symmetric:
\begin{equation}
\hat{G} (E) - \hat{G}_0 (E)  = \hat{G} (E) \hat{V} \hat{G}_0 (E) =  \hat{G}_0 (E)   \hat{V} \hat{G} (E).
\end{equation}

Hence, under reciprocity,  Eq.(\ref{OmegaGVG0}) is   given by
 \begin{equation}
  Tr \left [     \hat{ \Omega}_{E  }   \frac{d}{d E} \hat{\widetilde{\Omega}}^{\dag}_{E  }  - \hat{\widetilde{\Omega}}^{\dag}_{E  }   \frac{d}{d E}  \hat{ \Omega}_{E  }    \right ]     = 2  Tr \left [ \hat{G} (E) -  \hat{G}_0 (E)     \right ] , \label{OmegadiffG}
\end{equation}
where the trace on the right-hand side of Eq.(\ref{OmegadiffG}) in coordinate space is defined by
\begin{equation}
   Tr \left [ \hat{G} (E)     -  \hat{G}_0 (E)  \right ]  = \int_{-\infty}^\infty  d x  \langle x|  G ( E) -  \hat{G}_0 (E)  | x \rangle .
\end{equation}

In the end, Eq.(\ref{FriedelSGVG0})    result  a  Friedel  formula for complex local potential scattering systems: 
 \begin{align}
& \frac{1}{4 i}  Tr \left [ \widetilde{S}^\dag (k)   \frac{d}{d E}   S(k)  -  S(k)  \frac{d}{d E}   \widetilde{S}^\dag (k)   \right ] \nonumber \\
& = - Disc_E  \left [  \int_{-\infty}^\infty  d x  \langle x|     \hat{G} (E) -  \hat{G}_0 (E)    | x \rangle  \right ]  , \label{Friedelcomplexpot}
\end{align}
where  the hat on $S$-matrix operator in Eq.(\ref{FriedelSGVG0}) has been dropped, and now $S(k)/\widetilde{S}(k)$  in Eq.(\ref{Friedelcomplexpot})  represent the reduced on-energy-shell  $S$-matrix. Hence the trace on left-hand side of Eq.(\ref{Friedelcomplexpot}) is defined as regular trace of a matrix.    The discontinuity of Green's function crossing branch cut in complex $E$-plane  is defined by
\begin{equation}
Disc_E  \hat{G} (E)  = \frac{1}{2 i}  \left [  \hat{G} (E+ i 0 ) - \hat{G} (E- i 0)   \right ].
\end{equation}
We also remark that for a complex potential, the discontinuity of Green's function is not equivalent to the imaginary part of Green's function. This statement can be illustrated by considering the spectral representation of Green's function in Appendix \ref{spectralGreensec}.

Using unitarity relation of dual systems
$$ \widetilde{S}^\dag (k) =  S^{-1} (k),$$ and identity $$Tr \left [ \ln S(k ) \right ] = \ln \det \left [ S(k) \right ],$$
 the Friedel formula is thus also given in a more compact form
 \begin{align}
&  \frac{1}{2 i}         \frac{d}{d E}  \ln \left ( \det \left [   S(k)   \right ]  \right )   \nonumber \\
&=- Disc_E  \left [  \int_{-\infty}^\infty  d x  \langle x|     \hat{G} (E) -  \hat{G}_0 (E)    | x \rangle  \right ]  .  \label{DiscGlndetScompact}
\end{align}
Therefore, the Friedel formula is invariant under unitary transform of $S$-matrix, and doesn't depend on a specific basis of eigensolutions.

In a real potential scattering, the left-hand side of Friedel formula in Eq.(\ref{DiscGlndetScompact}) is real and positive, which is related to scattering phaseshift  matrix by
\begin{equation}
 \frac{1}{2 i}   \frac{d}{d E}  \ln \left ( \det \left [   S(k)   \right ]  \right ) =    \frac{d}{d E} Tr[\delta(k)].
\end{equation}
 However, for a complex potential in general, it is a complex matrix. As presented   in Appendix \ref{scattgen} and Appendix \ref{scattPTpot}, the $S$-matrix of complex potential dual systems cannot be parameterized by phaseshifts without additional symmetry constraints.

\subsubsection{Spatial inversion symmetry}
For a local and spatial inversion  ($\mathcal{P} $) symmetric potential, $$V(x)=V(-x),$$ as presented in Appendix \ref{Smatspatialinversion}, the $S$-matrix can be parameterized by two real phaseshifts, $\delta_\pm (k)$, and two real inelasticities, $\eta_\pm (k)$, see Eq.(\ref{Smatparityparameterization}):
   \begin{equation}
  S^{(+/-)}(k)   =  \begin{bmatrix}   e^{2 i \Delta_+ (k) }  &   0  \\   0 &  e^{2 i \Delta_- (k) }   \end{bmatrix}      ,
 \end{equation}
where
\begin{equation}
e^{2 i \Delta_\pm (k) }   = \eta_\pm (k) e^{2 i \delta_\pm (k) }  . \label{complexphaseshiftdef}
\end{equation}
 Hence,  the Friedel formula under $\mathcal{P} $ symmetry constraint  is determined by the sum of two complex functions, $\Delta_\pm (k)$, that play the role of complex phaseshifts,
   \begin{equation}
 \frac{1}{2 i}   \frac{d}{d E}  \ln \left ( \det \left [   S(k)   \right ]  \right ) =  \frac{d}{d E}  \left [ \Delta_+ (k)   +   \Delta_- (k)    \right ] . \label{QtimePsym}
\end{equation}

\subsubsection{$\mathcal{P} \mathcal{T}$ symmetry}
 For a local and $\mathcal{P} \mathcal{T}$ symmetric potential, $$V^*(x)=V(-x),$$   the $S$-matrix in parity basis can be parameterized by two real phaseshifts and one real inelasticity, see Eq.(\ref{PTSmatparity}) and Eq.(\ref{PTconstraint}). In parity basis, the  $S$-matrix is no longer diagonal, however, because of
\begin{equation}
\det  \left [ S^{(+/-)} (k )  \right ]  =   e^{2 i (\delta_+ (k)+\delta_- (k))  },
\end{equation}
the Friedel formula with $\mathcal{P} \mathcal{T}$ symmetry constraints doesn't depend on inelasticity, and is given by the sum of two real phaseshifts,
 \begin{equation}
 \frac{1}{2 i}   \frac{d}{d E}  \ln \left ( \det \left [   S(k)   \right ]  \right )   = \frac{d}{d E}    \left [  \delta_+ (k) +  \delta_-  (k)\right ]. \label{QtimePTsym}
\end{equation}
Using the expression of discontinuity of Green's function in Eq.(\ref{DiscGPTsym}),  Friedel  formula under  $\mathcal{P} \mathcal{T}$ symmetry  thus yields a real equation,
 \begin{align}
 & \frac{d}{d k}    \left [  \delta_+ (k) +  \delta_-  (k)\right ]     \nonumber \\
  &=         \sum_{p= \pm k}   Re \left [ \int_{-\infty}^\infty d x \Psi_{  k}  (x, p)  \Psi^*_{   k}  (-x, -p) \right ].
\end{align}

\section{Krein's theorem in complex potential scattering theory}\label{Kreintheorem}

\subsection{Krein's theorem in complex potential scattering theory and symmetry constraints} 

For a complex potential, using Friedel formula in Eq.(\ref{DiscGlndetScompact}), assuming both Green's function and $S$-matrix having the branch cuts along real   axis in complex $E$-plane,   the Green's function is thus constructed by Cauchy's integral through the discontinuity  of Green's function across both physical and unphysical cuts. Hence even with a complex potential, the expression of Krein's theorem given in Eq.(\ref{Kreinformula}) is still valid and remains the same.
We would also   point out that the integration of Green's function  over $x$ may bring down extra singularity factor, such as $1/k$. Hence in addition to the branch cut  that is inherited  from unintegrated Green's function itself, the integrated Green's function may have extra singularities, such as pole contribution at physical threshold because of the extra  $1/k$ factor brought down by integration. A simple example of singularities structure of integrated Green's function for the scattering with a complex contact interaction can be found in Sec.\ref{contactexample}.    Therefore, when extra singularity factors show up in integrated Green's function, though unintegrated Green's function still satisfies Cauchy's integral relation
\begin{equation}
\langle x|    \hat{G}(E)  | x \rangle = \frac{1}{\pi} \left [ \int_{-\infty}^{-E_L} +\int_0^\infty  \right ] d \epsilon \frac{Disc_\epsilon   \langle x|   \hat{G}(\epsilon)  | x \rangle }{\epsilon - E}, \label{dispG}
\end{equation}
the Cauchy's integral  of integrated Green's function in terms of discontinuity of integrated Green's function must be modified and pick up the contribution of extra singularities. The extra singularities contribution in integrated Green's function can also be understood by Krein's theorem in Eq.(\ref{Kreinformula}). Let's rewrite the right-hand side of  Eq.(\ref{Kreinformula}) by integration by parts, and also using Friedel formula in Eq.(\ref{DiscGlndetScompact}), we thus find
 \begin{align}
& \int_{-\infty}^\infty  d x  \langle x|     \hat{G} (E) -  \hat{G}_0 (E)    | x \rangle  =   - \frac{i}{2\pi }      \frac{  \ln \left ( \det \left [   S(  0)   \right ]  \right ) }{  E}   \nonumber \\
& + \frac{1}{\pi} \left [ \int_{-\infty}^{-E_L} +\int_0^\infty  \right ] d \epsilon \frac{ Disc_\epsilon \left [  \int_{-\infty}^\infty  d x  \langle x|     \hat{G} (\epsilon) -  \hat{G}_0 (\epsilon)    | x \rangle   \right ]  }{\epsilon- E} , \label{poledispersionintegratedG}
\end{align}
where the surface term on the right-hand side of above equation reflects the extra pole contribution of integrated Green's function, other surface terms are assumed vanishing and have been dropped.      The non-trivial  value of $\ln \left ( \det \left [   S(  0)   \right ]  \right )$ at physical threshold   thus determines the presence of  extra pole singularity of integrated Green's function.

 Using Eq.(\ref{QtimePsym}) and Eq.(\ref{QtimePTsym}) for $\mathcal{P}$ and $\mathcal{P} \mathcal{T}$ symmetric systems respectively, the integrated Green's functions are related to  phaseshifts  explicitly by
  \begin{align}
& \int_{-\infty}^\infty  d x  \langle x|     \hat{G} (E ) -  \hat{G}_0 (E)    | x \rangle  \nonumber \\
&=  -     \frac{1}{\pi} \left [ \int_{-\infty}^{-E_L} +\int_0^\infty  \right ] d \epsilon  \frac{   \Delta_+ (\sqrt{2 m \epsilon})  +  \Delta_- (\sqrt{2 m \epsilon}) }{(\epsilon- E)^2}   ,
\end{align}
for $\mathcal{P}$ symmetric systems,  and  
  \begin{align}
&  \int_{-\infty}^\infty  d x  \langle x|     \hat{G} (E ) -  \hat{G}_0 (E)    | x \rangle   \nonumber \\
&=  -    \frac{1}{\pi} \left [ \int_{-\infty}^{-E_L} +\int_0^\infty  \right ] d \epsilon  \frac{   \delta_+ (\sqrt{2 m \epsilon})  +  \delta_- (\sqrt{2 m \epsilon}) }{(\epsilon- E)^2}  ,
\end{align}
for $\mathcal{P} \mathcal{T}$ symmetric systems.

\subsection{Muskhelishvili-Omn\`es representation of Krein's theorem}

\subsubsection{Muskhelishvili-Omn\`es function}
In the nuclear/particle physics community,  the dispersion theoretical approach  by considering the analytical properties of reaction amplitude has been widely used in solving particles scattering, decaying and production problems, see e.g. Refs.~\cite{Gorchtein:2011vf,Danilkin:2014cra,Guo:2014vya,Guo:2015zqa,Guo:2016wsi}. The Muskhelishvili-Omn\`es (MO) representation \cite{muskhelishvili1941application,Omnes:1958hv}  that is sometimes also referred to $N/D$ method \cite{PhysRev.119.467,PhysRev.130.478} provides an elegant way of expressing the reaction amplitude as the product of two analytic functions: (1)  MO function that  possess only a right-hand branch cut  running along positive real $E$ axis, and the logarithm of  MO function  is given by Cauchy integral of phaseshift; (2) An analytic function that may possess other singularities but right-hand branch cut, such as left-hand branch cut, etc. The MO function is   constrained by unitarity relation that warrant the conservation of total probability of reactions.

Even in the complex potential scattering theory, MO dispersive approach still applies. However, we need to be aware that in general, especially with a complex potential, the discontinuity of a reaction amplitude is no longer the same as the imaginary part of the reaction amplitude, and the phaseshift function may become a complex function, etc.

Next,  we will give a brief and concise introduction to MO dispersive approach, so later on we can apply and cite the main result directly.  Let's consider a reaction amplitude, $F(E)$, which is analytic in complex $E$-plane and the property of $F(E)$ along the right-hand branch cut is constrained by unitarity relation,
\begin{equation}
e^{2 i  \Phi (E+ i0)} F(E-i 0) = F (E+ i 0), \ \ \ \ E \geqslant 0, \label{ReactionFampunitarity}
\end{equation}
where function  $\Phi (E\pm i0)$ that are defined above/below right-hand branch cut are related by
\begin{equation}
\Phi (E - i0) = - \Phi (E +  i0). \label{Phiphasecondition}
\end{equation}
The Eq.(\ref{ReactionFampunitarity}) defines the discontinuity of $F(E)$ across the right-hand branch cut,
\begin{equation}
Disc_{E } F(E) = \left [ \frac{e^{2 i \Phi^* (E +  i0) } -1}{2 i} \right ]^* F(E+ i0), \ \ \ \ E\geqslant 0.
\end{equation}
In the real potential scattering theory,   $\Phi (E)$ function  is real and   related directly to the elastic scattering phaseshift, see e.g. Refs.~\cite{muskhelishvili1941application,Omnes:1958hv}. Later on we will show that in $\mathcal{P}$ and $\mathcal{P} \mathcal{T}$ symmetric systems, $\Phi (E)$ is given by  sum of $\Delta_\pm (k)$ and $\delta_\pm (k)$ respectively, where condition of $\Phi(E)$ function in Eq.(\ref{Phiphasecondition}) is indeed satisfied. Eq.(\ref{ReactionFampunitarity}) suggests that the solution of $F(E)$ has the form of
\begin{equation}
F(E) = N(E) e^{ \lambda (E)},
\end{equation}
where $N(E)$ has no right-hand branch cut singularity,
\begin{equation}
N(E+ i0) = N(E- i 0), \ \ \ \ E \geqslant 0.
\end{equation}
The  $e^{ \lambda (E)}$ is usually referred as MO function or $D^{-1}(E)$ function. The  $ \lambda (E)$ has only right-hand branch cut singularity, and using Eq.(\ref{ReactionFampunitarity}), we find
\begin{equation}
Disc_E \lambda (E) = \Phi (E+ i 0),  \ \ \ \ E \geqslant 0,
\end{equation}
hence Cauchy integral theorem yields
\begin{equation}
\lambda (E) =  \frac{1}{\pi}  \int_0^\infty d \epsilon \frac{\Phi (\epsilon)}{\epsilon -E },
\end{equation}
and
\begin{equation}
F(E) = N(E) e^{ \frac{1}{\pi}  \int_0^\infty d \epsilon \frac{\Phi (\epsilon)}{\epsilon -E }}.
\end{equation}
The $N(E)$ usually describes the virtual physical processes, such as contributions  from $t$- and $u$-channel virtual particle exchange processes in a relativistic theory that produce a left-hand cut contribution below elastic threshold, see e.g. Ref.~\cite{PhysRev.123.692}.  The elastic scattering phaseshift is a physical observable, hence unitarity relation imposes a strong constraint on reaction amplitudes along right-hand branch cut. Unlike the unitarity relation above elastic threshold, the discontinuity of reaction amplitude across left-hand cut below elastic threshold in unphysical region  is normally less constrained and  heavily  model dependent. When the   left-hand cut is far away from the physical region, the $N(E)$ may be parameterized by approximate methods, such as conformal expansion \cite{Yndurain:2002ud} or simply treated as a constant \cite{Guo:2015zqa,Guo:2016wsi}. In non-relativistic potential scattering theory, for some short-range local potentials in 1D, such as contact interaction  or non-singular potential, it can be shown that $N(E)$ is indeed an energy independent constant, see the example in   Sec.\ref{contactexample}.

The argument of the right-hand cut solution can be extended into left-hand singularity as well. Assuming $\Phi (E)$ function is defined in both  physical region: $E\in [0, \infty]$, and unphysical region:  $E\in [-\infty, -E_L]$, so that Eq.(\ref{ReactionFampunitarity}) is now  valid for both right-hand and left-hand singularities. Both physical and unphysical branch cuts are now described by $\Phi (E)$ function.
In unphysical region:  $E\in [-\infty, -E_L]$, discontinuity of MO function vanish: 
$$e^{\lambda (E+ i0)} = e^{\lambda(E-i0)},$$
and the solution of $N(E)$ also has the form of, 
\begin{equation}
N(E) = N_0 e^{     \chi (E)},
\end{equation}
where $N_0$ is a normalization constant, and
\begin{equation}
Disc_E   \chi (E) = \Phi (E+ i 0),  \ \ \ \ E  \in [- \infty, -E_L].
\end{equation}
Therefore Cauchy integral theorem yields
\begin{equation}
 \chi (E) =  \frac{1}{\pi}  \int_{-\infty}^{-E_L} d \epsilon \frac{\Phi (\epsilon)}{\epsilon -E }. 
\end{equation}
With both left-hand and right-hand singularities described by $\Phi (E)$ function, thus, we finally get
\begin{equation}
F(E) = N_0 e^{ \frac{1}{\pi}  \left [  \int_{-\infty}^{-E_L} + \int_0^\infty \right ] d \epsilon \frac{\Phi (\epsilon)}{\epsilon -E }}. \label{leftrightcutsolution}
\end{equation}

\subsubsection{Muskhelishvili-Omn\`es representation of Krein's theorem in $\mathcal{P}  $ symmetric systems}
For a $\mathcal{P}  $ symmetric system, using unitary transform relation in Eq.(\ref{Sparityplanetransform}), the transmission amplitude $t(k)$ is related to $\Delta_\pm (k)$ by
\begin{equation}
 t(k)  = \frac{ e^{2 i \Delta_+ (k) }  + e^{2 i \Delta_- (k) } }{2} .
\end{equation}
The unitarity relation constraint in Eq.(\ref{extraconstraints}) yields
  \begin{equation}
     \Delta_\pm (-k)  = - \Delta_\pm (k)        , \label{DelatphaseconditionPsym}
 \end{equation}
hence we   find
  \begin{equation}
  e^{ 2 i \left [ \Delta_+ (k) + \Delta_- (k)  \right ] }  t(-k)  = t(k) . \label{transmissionampunitarity}
 \end{equation}
 Eq.(\ref{transmissionampunitarity}) and Eq.(\ref{DelatphaseconditionPsym}) are exact MO representation type in Eq.(\ref{ReactionFampunitarity}) and Eq.(\ref{Phiphasecondition}):
 \begin{equation}
 t(\pm k) = F(E \pm i 0), \ \ \ \  \Delta_+ (\pm k) + \Delta_- ( \pm k)  = \Phi(E\pm i0).
 \end{equation}
Therefore, MO representation of  transmission amplitude, $t(k)$, is given by
   \begin{equation}
\ln \left [ \frac{ t(k) }{N_0} \right ] =  \frac{1}{\pi}  \left [ \int_{-\infty}^{-E_L}  +  \int_0^\infty  \right ]  d \epsilon \frac{ \Delta_+ (\sqrt{2 m \epsilon}) + \Delta_- (\sqrt{2 m \epsilon})  }{\epsilon - E - i 0},
 \end{equation}
 and    Krein's theorem can also be written as
   \begin{equation}
-    \frac{d}{d E} \ln  t(k)   =  \int_{-\infty}^\infty  d x  \langle x|     \hat{G} (E+ i 0) -  \hat{G}_0 (E+ i 0)    | x \rangle   . \label{MOrepKreintheorem}
\end{equation}

\subsubsection{Muskhelishvili-Omn\`es representation of Krein's theorem in $\mathcal{P}  \mathcal{T}$ symmetric systems}
Similarly, in  $\mathcal{P}  \mathcal{T} $ symmetric system,  using expression of  the transmission amplitude $t(k)$ in Eq.(\ref{tandrampsPTsym}) and unitarity constraint in Eq.(\ref{extraconstraintsonPTsym}): $ t(-k ) = t^* (k)$,  we find
\begin{equation}
\eta (-k)  = \eta (k), \ \ \ \   \delta_\pm (-k)  = -\delta_\pm (k) ,
\end{equation}
and
\begin{equation}
 e^{ 2 i \left [ \delta_+  (k) +  \delta_- (k) \right ] }  t(- k)   = t(k).
\end{equation}
Hence, in $\mathcal{P}  \mathcal{T} $ symmetric system, 
 \begin{equation}
 t(\pm k) = F(E \pm i 0), \ \ \ \  \delta_+ (\pm k) + \delta_- ( \pm k)  = \Phi(E\pm i0),
 \end{equation}
 and MO representation of  Krein's theorem  has the same form as in Eq.(\ref{MOrepKreintheorem}).

\section{Discussion and summary}\label{summary}

Before we summarize results of our finding, a simple and   exactly solvable example of a particle scattering with a contact interaction is presented in below, which is sufficient to demonstrate numbers of unique features of complex scattering theory, such as spectral singularities etc.

\subsection{A simple example of particle scattering with a complex contact potential}\label{contactexample}  

\subsubsection{Scattering solutions}
Let's consider a simple but intuitive example of scattering solutions with a complex contact potential, 
\begin{equation}
V(x) = V \delta(x),  \ \ \ \ V = |V| e^{ i \theta}.
\end{equation}
The scattering solutions can be obtained rather straightforwardly by considering Eq.(\ref{LSabsorb}) and Eq.(\ref{LSemissive}), hence for an absorbing system, we find
  \begin{equation}
  \Psi_{  k} (x,p)  = e^{     i p x}   + i f_k e^{i k |x|},  
 \end{equation}
 where on-shell amplitude   depends only on $k$ as the result of contact interaction,
  \begin{equation}
   f_k  =-  \frac{m V}{k + i m V}.    
 \end{equation}
 For an emissive system, we thus have
   \begin{equation}
\widetilde{  \Psi}_{  k} (x,p) = e^{     i p x}   + i \widetilde{f}_k   e^{i k |x|}, 
 \end{equation}
 where
   \begin{equation}
  \widetilde{ f}_k  =-  \frac{m V^* }{k + i m V^*} = - f^*_{-k} .  
 \end{equation}
  
 As a symmetric potential, only one transmission and reflection amplitudes are needed,
 \begin{equation}
 t(k) = 1+ if_k , \ \ \ \ r(k) = i f_k ,
 \end{equation}
 the $S$-matrix in parity basis is thus given by
    \begin{equation}
  S^{(+/-)}(k)   =  \begin{bmatrix}  1+2 if_k &   0  \\   0 &  1 \end{bmatrix}   =  \begin{bmatrix}   \frac{k - i m V}{k + i m V} &   0  \\   0 &  1 \end{bmatrix}       .
 \end{equation}
 For a contact interaction, only positive parity solution survives. Similarly for an emissive system, we obtain
    \begin{equation}
  \widetilde{S}^{(+/-)}(k)   =  \begin{bmatrix}  1+2 i \widetilde{f}_k &   0  \\   0 &  1 \end{bmatrix}   =  \begin{bmatrix}   \frac{k - i m V^* }{k + i m V^*} &   0  \\   0 &  1 \end{bmatrix}       .
 \end{equation}
 Hence, the unitarity is indeed given by
   \begin{equation}
  \widetilde{S}^{(+/-) \dag }(k) S^{(+/-)}(k)     =  \begin{bmatrix}   \frac{k + i m V }{k - i m V} &   0  \\   0 &  1 \end{bmatrix}   \begin{bmatrix}   \frac{k - i m V}{k + i m V} &   0  \\   0 &  1 \end{bmatrix}   = \mathbb{I}    .
 \end{equation}

 The complex functions, $\Delta_\pm (k)$, are given by
  \begin{equation}
 \Delta_+ (k) =  - \cot^{-1} \left ( \frac{k}{m V} \right ), \ \ \ \   \Delta_- (k) =0, \label{Deltaphasedeltapot}
 \end{equation}
 and  
 \begin{equation}
  \frac{d }{dE}  \Delta_+ (k) =  \frac{m}{k} \frac{m V}{k^2 +(mV)^2 }.
 \end{equation}

\subsubsection{Green's function solution}

The Dyson equation for a contact potential is given by an algebra equation,
\begin{equation}
G(x,x';E) = G_0(x-x';E) +  G_0(x;E) V G(0,x';E) ,
\end{equation}
where $G_0(x;E)$ is defined in Eq.(\ref{G0def}).
The solution of Green's function is thus given analytically by
\begin{equation}
G(x,x';E+ i 0) =  - \frac{i m}{k} \left [ e^{i k |x- x'|}  - \frac{i m V}{k+ i m V} e^{i k ( |x| +| x'|)}  \right ]. \label{Gsoldeltapot}
\end{equation}

\subsubsection{Friedel formula check}

The Green's functions that is defined below real $E$ axis, $E- i0$, are simply obtained by replacing $k$ by $- k$ in above expressions. Thus  we find
\begin{equation}
 -Disc_E \int_{-\infty}^\infty d x \left [ G(x,x;E) - G_0(0;E)   \right ] =  \frac{m}{k}  \frac{m V}{k^2 +(mV)^2 }.
\end{equation}
The discontinuity of integrated Green's function difference  is  hence a complex function as well and indeed equal to $\frac{d }{dE}  \Delta_+ (k) $. The left-hand branch cut is absent for a contact interaction.  The Friedel formula is   satisfied.

\subsubsection{Krein's theorem check}

Using analytic expression given in Eq.(\ref{Gsoldeltapot}) and Eq.(\ref{Deltaphasedeltapot}), thus  we can show easily that Krein's theorem is indeed also satisfied for a contact interaction,
\begin{align}
& \int_{-\infty}^\infty d x \left [ G(x,x;E + i0 ) - G_0(0;E + i 0)   \right ]  \nonumber \\
& = -  \frac{1}{\pi} \int_0^\infty d \epsilon  \frac{   \Delta_+ (\sqrt{2 m \epsilon})    }{(\epsilon- E-i0)^2}  =- \frac{m}{k^2} \frac{i m V}{k+ i mV}.
\end{align}
From above expression, we can see clearly that the presence of a pole contribution $1/k^2$ in addition to the branch cut singularity, and $\Delta_+ (0) = - \frac{\pi}{2}  $ is indeed non-zero. For $\theta \in [0, \frac{\pi}{2}]$, the interaction is repulsive-like, we find
\begin{align}
&  \frac{1}{\pi} \int_0^\infty d \epsilon  \frac{      Disc_\epsilon \int_{-\infty}^\infty d x \left [ G(x,x;\epsilon) - G_0(0;\epsilon)   \right ]   }{\epsilon- E-i0}    \nonumber \\
&  =   \frac{m}{k} \frac{1}{k + im V}, \label{DispGcontactpot}
\end{align}
hence we can   verify that the Cauchy integral equation for integrated Green's function is indeed Eq.(\ref{poledispersionintegratedG}) type,
\begin{align}
& -     \frac{  m     }{  k^2 }  +  \frac{1}{\pi} \int_0^\infty d \epsilon  \frac{      Disc_\epsilon \int_{-\infty}^\infty d x \left [ G(x,x;\epsilon) - G_0(0;\epsilon)   \right ]   }{\epsilon- E-i0}   \nonumber \\
&  =   \int_{-\infty}^\infty d x \left [ G(x,x;E+ i0) - G_0(0;E+ i0)   \right ].  \label{Cauchyintegralrepcontactpot}
\end{align}

For the contact interaction, the MO representation of transmission amplitude, $t(k)$, only has  a physical branch cut singularity, 
\begin{equation}
t(k) = \frac{k}{k+ i m V}  =N_0 e^{ \frac{1}{\pi} \int_0^\infty d \epsilon  \frac{   \Delta_+ (\sqrt{2 m \epsilon})    }{\epsilon- E-i0} },
\end{equation} 
where  $N_0$ is a constant and simply plays the role of integral subtraction to ensure the fast convergence of dispersive integral,
\begin{equation}
N_0 = t(i \kappa_0 ) e^{ -\frac{1}{\pi} \int_0^\infty d \epsilon  \frac{   \Delta_+ (\sqrt{2 m \epsilon})    }{\epsilon + \frac{\kappa_0^2}{2 m}} },
\end{equation}
and $\kappa_0$ can be chosen arbitrarily. Hence, the  MO representation of Krein's theorem is indeed given by Eq.(\ref{MOrepKreintheorem}).

\subsubsection{Spectral singularity and bound state above physical threshold}
It has been well-known that in non-Hermitian complex potential scattering theory, the bound state may appear above a physical threshold, which is usually referred to spectral singularities \cite{PhysRevLett.102.220402,Ahmed_2009,PhysRevB.80.165125}. It was shown in Ref.~\cite{PhysRevLett.102.220402} that spectral singularities of a non-Hermitian Hamiltonian yields   divergences of reflection and transmission coefficients of scattered states,  and are interpreted as resonance states with vanishing spectral width. The origin of zero-width resonances and bound states are nevertheless the same, both are the results of pole solutions in dynamical related amplitudes and quantities, such as scattering amplitudes and Green's function, etc. Conventionally the  pole solutions below physical threshold are referred as bound states, contrast to the spectral singularity related zero-width resonances in non-Hermitian scattering theory that appear in physical continuous spectrum. This can be easily understood by the simple example of contact interaction scattering:  the   pole singularity of dynamical system is proportional to $\frac{1}{k+i mV}$,  hence for the real potential scattering, the pole solution, $k_{pole} = - i m V $, correspond to  a bound state for attractive potential ($V<0$) or a virtual bound state if potential is repulsive ($V>0$). On complex $E$-plane,  the bound state solution is located on the first Riemann sheet  below physical threshold,  $2 m E_B= - (mV)^2$, and the  virtual bound state however is   on  the second Riemann sheet   (unphysical sheet). In the complex potential scattering, $V= |V| e^{i \theta}$   has the access to the entire complex plane, hence when $\theta$ is rotated from $0$ to $\pi$, now the pole solution can move from  below the physical threshold on the first Riemann sheet into the second Riemann sheet by crossing the positive real $E$ axis from below. Therefore the spectral singularity of zero-width resonance occurs, $2 m E_{pole} = (m |V|)^2$, at $\theta = \frac{\pi}{2}$.

  \begin{figure}
\begin{center}
\includegraphics[width=0.8\textwidth]{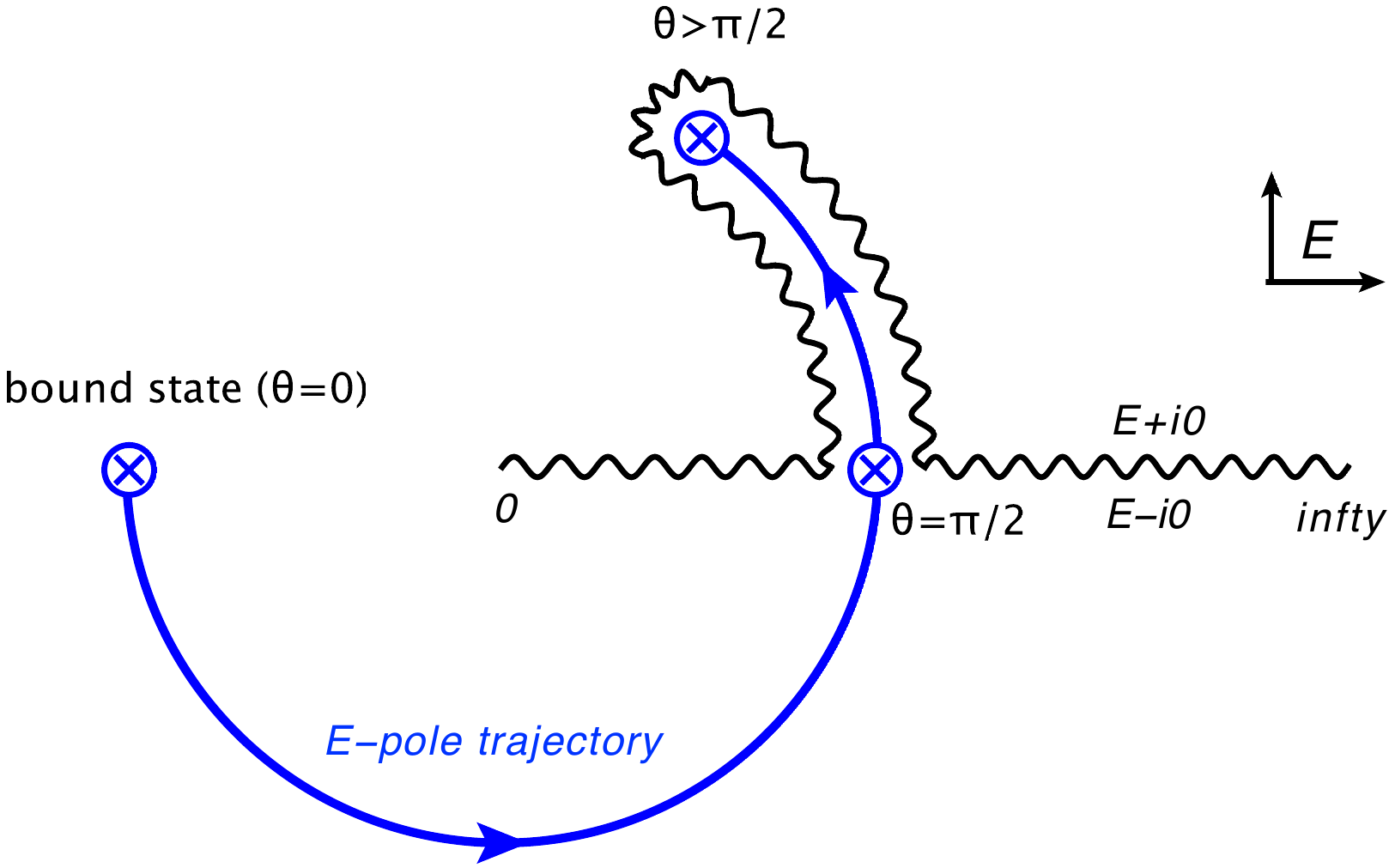}
\caption{    The plot of motion of pole singularity on the first Riemann sheet of complex $E$-plane, the pole position is given by $E_{pole} =- (m |V|)^2 e^{2 i \theta}$ as $\theta$ increases from $0$ upto $[\frac{\pi}{2}, \pi]$ region. The trajectory of motion of pole is represented by solid blue curve.}\label{poleplot}
\end{center}
\end{figure}

The motion of pole singularities in complex plane also affect the Cauchy integral representation of Green's function, the    Eq.(\ref{dispG}) is valid only when $\theta \in [0, \frac{\pi}{2}] $, the pole is located right below contour of integration over positive real $E$ axis on physical sheet. When the value of $\theta$ is further increased into  $[\frac{\pi}{2}, \pi]$ region,   the pole starts moving across integral contour into the second Riemann sheet.  The motion of pole hence drags the contour of integration moving with it to keep Cauchy integral well-defined on physical sheet, see Fig.~\ref{poleplot}. In the end, the extra term as the residue of deformed contour of integration must be added into  Cauchy integral representation of Green's function. The same   is true for Cauchy integral representation of integrated Green's function in Eq.(\ref{DispGcontactpot}) and Eq.(\ref{Cauchyintegralrepcontactpot}), it is sufficient to demonstrate spectral singularity  by considering expression in Eq.(\ref{DispGcontactpot})
\begin{align}
&   \frac{1}{\pi} \int_0^\infty d \epsilon  \frac{      Disc_\epsilon \int_{-\infty}^\infty d x \left [ G(x,x;\epsilon) - G_0(0;\epsilon)   \right ]   }{\epsilon- E}   \nonumber \\
&= -\frac{1}{\pi} \int_0^\infty d \epsilon \frac{1}{\epsilon - E}  \frac{m}{ \sqrt{2m \epsilon }}  \frac{m V}{2 m \epsilon +(mV)^2 }   .
\end{align}
For $\theta \in [0, \frac{\pi}{2}]$, we already know the result of integration is  
\begin{equation}
 -\frac{1}{\pi} \int_0^\infty d \epsilon \frac{1}{\epsilon - E -i 0}  \frac{m}{ \sqrt{2m \epsilon }}  \frac{m V}{2 m \epsilon +(mV)^2 } = \frac{m}{k} \frac{1}{k + im V}  .
\end{equation}
The pole singularity that is determined by condition $$2 m \epsilon +(mV)^2 =0$$ is now well illustrated on the left-hand side of above equation.   When $\theta $ value is increased from $\theta \in [0, \frac{\pi}{2}]$ into $\theta \in [ \frac{\pi}{2}, \pi]$, the pole moves from below contour of integral into above,   hence the contour of integral must be deformed to follow the motion of pole, see Fig.~\ref{poleplot}, so that the residue contribution due to the deformation of integration contour must be added,    we find
\begin{align}
&    \frac{1}{\pi} \int_0^\infty d \epsilon  \frac{      Disc_\epsilon \int_{-\infty}^\infty d x \left [ G(x,x;\epsilon) - G_0(0;\epsilon)   \right ]   }{\epsilon- E-i0}     \nonumber \\
&+ \frac{2  m }{ k^2 + (mV)^2  } = \frac{m}{k}  \frac{1}{k+ im V} , \ \ \ \  \theta \in [\frac{\pi}{2}, \pi].
\end{align}

\subsection{Summary and outlook}

In summary,  as the consequence of the balanced gain and loss dual systems, the biorthogonal relation can be established between eigenstates of dual systems. Hence the Friedel formula in complex potential scattering theory is still given in the  same form,
 $$  \frac{1}{2 i}         \frac{d}{d E}  \ln \left ( \det \left [   S(k)   \right ]  \right ) =- Disc_E \left [  Tr\left [     \hat{G} (E) -  \hat{G}_0 (E)     \right ]  \right ] ,  $$
the same is true for Krein's theorem given   in Eq.(\ref{Kreinformula}). The same mathematical forms of Friedel formula and Krein's theorem in both real and complex potential scattering theories suggest that  Friedel formula and Krein's theorem for  real and complex potentials can be simply connected by analytical continuation. This argument may be also supported by Muskhelishvili-Omn\`es representation of Krein's theorem in  Eq.(\ref{MOrepKreintheorem}). Therefore, numbers of useful relations in real potential scattering theory may still apply to complex systems, such as $\mathcal{P} \mathcal{T}$ symmetric systems.

One of these useful relations that is closely related to   Eq.(\ref{MOrepKreintheorem})   is the formula given in  Refs.~\cite{PhysRevB.51.6743,PhysRevA.54.4022},
\begin{align}
& - \frac{d 
\ln t (k)}{d E }   - \frac{  r^{(L)}(k) +r^{(R)}(k) }{4 E}e^{2ikL}    \nonumber \\
&=\int_{-L}^{L} dx  \langle x|  \hat{G}( E + i0 ) - \hat{G}_0( E+i0) | x \rangle. \label{eq:travx}
\end{align}
The Eq.(\ref{eq:travx}) shows the relation between   partially integrated Green's function up to a range $L$ and  both transmission   $t(k)$  and reflection $r^{(R/L)}(k)$  amplitudes for   a finite range  potential scattering system.  The potential  regardless the specific shape   may be  approximated by the sum of multiple-layers of square well potential.   After some lengthy derivations, see  Refs.~\cite{PhysRevB.51.6743,PhysRevA.54.4022},   Eq.(\ref{eq:travx}) can be obtained remarkably. As 
pointed out in Ref.~\cite{PhysRevA.54.4022}, a calculation of the density of states without taking into account the extra oscillation term in Eq.(\ref{eq:travx}) yields a wrong result. Such oscillations in density of states and the partial density of states 
  influence the conduction properties of sufficiently small conductors \cite{PhysRevB.40.3409}. At the limit of $ L \gg 1/k$, or in  cases such as the resonant scattering where reflection is negligible,  
the second oscillation term in Eq.(\ref{eq:travx}) can be neglected, and Eq.(\ref{MOrepKreintheorem})  is recovered. The relations given in   Eq.(\ref{eq:travx}) and Eq.(\ref{MOrepKreintheorem})  may also be valid in   describing the dynamics of  other waves, such as acoustic or electromagnetic waves, as far as its propagation in the medium is  a
second-order Schr\"odinger equation-like  differential equation \cite{Aronov_1991}. For a example,  in Ref.~\cite{PhysRevLett.75.2312}, 
a similar result   to  relation in Eq.(\ref{eq:travx}) is obtained for an electromagnetic wave propagating in a finite system with an arbitrary position-dependent refractive index that plays the role of interaction potential. The partially integrated Green's function over the finite range of the scattering region naturally appears in the theory of calculating the expectation value of the spin components along or perpendicular to the direction of the external magnetic field. In addition,
it also arise  in the general analysis of so called the B{\"u}ttiker–Landauer tunneling time through a real potential.
The question how the average value of the spin or tunneling time components behave explicitly in the case of   $\mathcal{P} \mathcal{T}$ symmetric systems has  not been properly studied.
  On top of the above mentioned cases,  similar  expression  also arise for “cooperon” in the theory of weak localization and weak anti-localization in semiconductor films.

One of the important features of the Friedel formula in complex potential scattering theory is that the absorptive part of Green's function is in general a complex function and  no longer related to the conventional definition of density of states   of Hermitian quantum theory.  In  the case of $\mathcal{P} \mathcal{T}$ symmetric systems, the   absorptive part of Green's function is  real, hence, the integrated absorptive part of Green's function may still be interpreted as  time delay function for dual systems with balanced gain and loss. The  imaginary part of Green's function in  $\mathcal{P} \mathcal{T}$ symmetric systems may be considered as generalized density of states,    it is still a conserved quantity but no longer positive-definite. Hence it is also referred as pseudo-norm    in Refs.~\cite{doi:10.1142/S0217732301005333,LEVAI2002271}. An alternative view of physical interpretation of biorthogonal quantum theory is given in Ref.~\cite{Brody_2013}:  the dual Hilbert spaces where the dual systems are defined in are replaced by a single Hilbert space with a non-trivial metric     that connect eigenstates of dual systems, hence the  physical observable    is thus evaluated as  the expectation value   in the Hilbert space endowed with a non-trivial metric. Similarly, the absorptive part of Green's function in  complex potential scattering theory now may be interpreted as the density of states in the Hilbert space with a non-trivial metric that  describes the absorbing/emissive nature of  the complex system.

There are a number of problems  in  $\mathcal{P}\mathcal{T}$-symmetric systems that are worth studying in detail. For instance, finite size effect and related Tamm states. These Tamm states
arising at the boundary of any finite semiconductor are practically independent of the distribution of defects and external perturbations. Another problem that is less discussed in the literature for a $\mathcal{P} \mathcal{T}$ system is the field dependence of the anomalous magnetoresistance for a sample with a thickness at the order of the external magnetic length. In such a case, the interference effects associated with the boundary become very important,  affect the charge's quantum transport and may lead to interesting oscillations of the magnetoresistance.
 
\acknowledgments

We   acknowledge support from the Department of Physics and Engineering, California State University, Bakersfield, CA. We also thanks Christopher Wisehart  for improving the use of the English language in the manuscript.

\appendix

\section{Scattering theory for a complex potential}\label{scattgen}

In this section, we give a brief description of scattering theory for a complex potential in general in one dimensional space, good references can be found in  Refs.~\cite{FESHBACH1985398,MUGA2004357,Brody_2013}.

\subsection{Scattering solutions of an absorbing system} 
 In terms of Lippmann-Schwinger (LS) equation, the wavefunction of an absorbing system that is defined above the real axis in a complex $E$-plane,
  \begin{equation}
  \Psi_{  k}  (x)  = \langle x | \Psi_{E+ i 0} \rangle , \ \ \ \ k= \sqrt{2 m (E+ i 0)},
 \end{equation}
 satisfies integral equation
  \begin{align}
 & \Psi_{  k} (x, p)= e^{     i p x}  \nonumber \\
 & +  \int_{-\infty}^\infty  d x'  G_0 (x-x'; E+ i 0) V(x')  \Psi_{  k}  (x',p) , \label{LSabsorb}
 \end{align}
 where the symbol  $$ p = \pm k$$ is used to label  two independent boundary conditions:   right ($e^{  i k x}$) and left ($e^{ - i k x}$) propagating   incoming plane waves respectively.
 The Green's function of free propagating particle is given by
 \begin{equation}
 G_0 (x; E+ i 0) = \int_{-\infty}^\infty \frac{d p}{2\pi} \frac{e^{i p x}}{E - \frac{p^2}{2m} + i 0} = -   \frac{  i m }{ k}   e^{i k |x|} . \label{G0def}
 \end{equation}
 The on-shell scattering amplitudes for an absorbing system is thus defined through the asymptotic behavior of wavefunction,
 \begin{equation}
  \Psi_{  k} (x,p) \stackrel{|x| \geqslant L}{\rightarrow} e^{     i p x}   + i f_k( p' ,   p ) e^{i k |x|}, \ \  \ \ p' = k\frac{x}{|x|}, \label{asymwavabsorb}
 \end{equation}
 where $L$ stands for the range of potential: $V(x) \stackrel{|x| \geqslant L}{\rightarrow}  0$.  The on-shell  scattering amplitudes with right/left propagating waves are given by
  \begin{equation}
   f_k ( p' ,    p )  =- \frac{m}{k} \int_{-\infty}^\infty  d x'  e^{- i p' x'} V(x')  \Psi_{  k}  (x', p) ,   \label{scatampabsorb}
 \end{equation}
 where $$(p',p) \in \pm k.$$
After removing the $\delta$-functions that preserve the energy conservation between initial and final scattering states, the reduced $S$-matrix for an absorbing system in right/left propagating plane wave basis is thus defined by
 \begin{equation}
 S^{(R/L)}(k) = \begin{bmatrix} t^{(R)} (k) & r^{(L)} (k) \\  r^{(R)} (k) & t^{(L)} (k)  \end{bmatrix}   , \label{SRLmatabsorb}
 \end{equation}
 where $ t^{(R/L)}   $ and $ r^{(R/L)}  $ denote the transmission and reflection amplitudes, superscripts $(R/L)$ are adopted to label amplitudes with boundary condition of   right/left propagating waves respectively.  The transmission and reflection amplitudes are related to  scattering amplitudes by
  \begin{equation}
  \begin{bmatrix} t^{(R)} (k) & r^{(L)} (k) \\  r^{(R)} (k) & t^{(L)} (k)  \end{bmatrix}  = \begin{bmatrix} 1+ i f_k (   k,   k) &  i f_k (  k,   -k) \\   i f_k ( - k,   k)& 1+ i f_k ( - k, - k) \end{bmatrix}   . \label{trscattamp}
 \end{equation}

\subsection{Scattering solutions of an emissive system} 
 Similarly, for an emissive system with a complex potential $V^*(x)$, the wavefunction solution that is  defined above the real axis in a complex $E$-plane
   \begin{equation}
\widetilde{  \Psi}_{   k}  (x)  = \langle  x | \widetilde{\Psi}_{E+ i 0} \rangle  ,
 \end{equation}
  is also given by LS equation,
  \begin{align}
 & \widetilde{\Psi}_{  k} (x,p)= e^{     i p x}  \nonumber \\
 & +  \int_{-\infty}^\infty  d x'  G_0 (x-x'; E+ i 0) V^*(x') \widetilde{ \Psi}_{  k}  (x', p) .   \label{LSemissive}
 \end{align}
    Hence the  on-shell  scattering amplitudes for an emissive system is defined by
  \begin{equation}
  \widetilde{ f}_k( p' ,    p )  =- \frac{m}{k} \int_{-\infty}^\infty  d x'  e^{- i p' x'} V^*(x')  \widetilde{\Psi}_{  k}  (x',p) , \label{scatampregen}
 \end{equation}
 and
  \begin{equation}
\widetilde{  \Psi}_{  k} (x,p) \stackrel{|x| \geqslant b}{\rightarrow} e^{     i p x}   + i \widetilde{f}_k ( p' ,   p ) e^{i k |x|}, \ \  \ \ p' = k\frac{x}{|x|}.
 \end{equation}
 The $S$-matrix  for an emissive system in right/left propagating plane wave basis is thus given by
 \begin{equation}
 \widetilde{S}^{(R/L)}(k) = \begin{bmatrix} \widetilde{t}^{(R)} (k) & \widetilde{r}^{(L)} (k) \\  \widetilde{r}^{(R)} (k) & \widetilde{t}^{(L)} (k)  \end{bmatrix}  , \label{SRLmatemissive}
 \end{equation}
   where $ \widetilde{t}^{(R/L)}   $ and $\widetilde{ r}^{(R/L)}  $ are the transmission and reflection amplitudes for an emissive system,  and they are related to $  \widetilde{ f}_k ( p' ,    p )$ in a similar way as   in Eq.(\ref{trscattamp}).

\subsection{The relations of wavefunctions and scattering amplitudes in dual systems} 
First of all, the complex conjugate of Eq.(\ref{LSemissive}) for an emissive system yields
    \begin{align}
 & \widetilde{\Psi}^{  *}_{  k} (x, p) \nonumber \\
 &= e^{     i (- p) x}    - \frac{i m}{ (-k)}  \int_{-\infty}^\infty  d x'  e^{i (- k) |x-x'|}  V(x') \widetilde{ \Psi}^{  *}_{  k}  (x', p) ,  
 \end{align}
 where $$k =\sqrt{2 m (E + i 0)}$$ and $ \widetilde{\Psi}^{  *}_{  k} (x, p) $ is defined above real axis in complex $E$-plane.  Compared with LS equation of an absorbing system that is defined below real axis in complex $E$-plane,
    \begin{align}
&  \Psi_{  -k} (x, -p) \nonumber \\
&= e^{     i (-p) x}  - \frac{i m}{(-k)}  \int_{-\infty}^\infty  d x'  e^{i (-k) |x-x'|} V(x')  \Psi_{-  k}  (x', - p) ,
 \end{align}
  where $$-k =\sqrt{2 m (E - i 0)},$$ we conclude that the wavefunctions of dual systems are related by
    \begin{equation}
  \widetilde{\Psi}^{ *}_{  k} (x, p)=  \Psi_{    -k } (x, -p) , \ \ \ \ p = \pm k. \label{wavdualproj}
 \end{equation}
 This relation  in fact is the explicit expression of Eq.(\ref{wavdualrel}) after the projection of state operator into position space and also taking into account the boundary conditions.
   
 Next, using  the definition of scattering amplitudes of dual systems in Eq.(\ref{scatampabsorb}) and Eq.(\ref{scatampregen}),  combined with relation of wavefunctions in Eq.(\ref{wavdualproj}),    we find that the on-shell scattering amplitudes of dual systems are  related by
      \begin{equation}
  \widetilde{f}^{*}_k ( p' ,   p )  =- f_{- k} ( -p' ,   -p ), \ \ \ \   (p', p) \in \pm k . \label{scattampdual}
 \end{equation}

\subsection{Unitarity relation of $S$-matrix of dual systems} 

\subsubsection{Unitarity relation of dual systems}
The unitarity relation of dual systems is given by
    \begin{align}
 \widetilde{S}^{(R/L)\dag}(k)   S^{(R/L)}(k)&=  \mathbb{I}   . \label{unitarityrelation}
 \end{align}
 As shown in Eq.(\ref{SRLmatabsorb}) and Eq.(\ref{SRLmatemissive}), in general the $S$-matrix for both an absorbing system and an emissive system dependents on four independent complex functions.  Superficially eight complex functions are required to describe the dynamics of dual systems, unitarity relation provides four complex constraint equations on eight dynamical functions. Hence, the $S$-matrix of an emissive system is determined completely by $S$-matrix of absorbing system,
   \begin{equation}
   \widetilde{S}^{(R/L)\dag} (k)  = \left [ S^{(R/L) }( k) \right ]^{-1}.
 \end{equation}
  In the end, for a general non-symmetric complex potential, four independent complex dynamical functions are required. Using Eq.(\ref{scattampdual}), we also find,
 \begin{equation}
   \widetilde{S}^{(R/L)\dag} (k)  = \left [ S^{(R/L) }( -k) \right ]^T . 
 \end{equation}
 We   remark that though in present work, the terminology "unitarity relation" is constantly used to describe the relation in Eq.(\ref{unitarityrelation}),  we must be aware that in complex potential theory, Eq.(\ref{unitarityrelation}) only refers to balanced gain and loss between dual systems instead of  probability preserving unitary time evolution     in  Hermitian scattering theory.

\subsubsection{$S$-matrix in parity basis} 
 For many occasions, especially in the cases that the potential displays spacial reflection symmetries,  it is more convenient to use the scattering solutions with boundary conditions of positive parity  ($\cos kx$) and negative parity  ($i \sin k x$) propagating incoming waves. The wavefunction and scattering amplitude solutions with different boundary conditions are related simply by the linear superposition:
 \begin{equation}
   \Psi^{(+/-)}_{  k} (x ) = \frac{ \Psi_{  k} (x, k) \pm  \Psi_{  k} (x, -k)}{2}    , \label{wavparitydef}
 \end{equation}
 and
    \begin{equation}
    f^{(+/-)}_k ( p'  )   = \frac{ f_k ( p' ,    k )   \pm  f_k ( p' ,   - k )  }{2}    , \ \ \ \ p'=\pm k, \label{ampparitydef}
 \end{equation}
where the superscripts $(+/-) $ are used to label solutions that  correspond respectively to positive/negative parity  propagating incoming waves: $\cos kx/i \sin k x$. The symbol $p=\pm k$ that is used to label  solutions that correspond to right/left propagating incoming waves  is  hence dropped and replaced by labels: $(+/-) $.

 The $S$-matrix in different bases are related by a unitary transformation,
    \begin{equation}
  S^{(+/-)}(k) =U^\dag S^{(R/L) } (k) U      , \label{Sparityplanetransform}
 \end{equation}
 where  $S^{(+/-)}(k) $ stands for the $S$-matrix in parity basis, and $U$-matrix is given by
   \begin{equation}
  U  =  \begin{bmatrix} \frac{1}{\sqrt{2}}& \frac{1}{\sqrt{2}}\\  \frac{1}{\sqrt{2}} & - \frac{1}{\sqrt{2}} \end{bmatrix}   .
 \end{equation}

\section{Symmetry constraints in complex potential scattering theory}\label{scattPTpot}

\subsection{Reciprocity and spatial inversion symmetry}
In addition to the unitarity constraints, the symmetries of potential also impose extra constraints on dynamical system and further reduce the number of independent complex functions in describing dynamical systems. Most commonly considered symmetries are time reversal $(\mathcal{T})$, spatial inversion $(\mathcal{P})$, reciprocity $(\mathcal{R})$, and combined  $\mathcal{P} \mathcal{T}$ symmetry. The time reversal symmetry holds when $\hat{V} = \hat{V}^*$ is satisfied, hence for systems with  complex potentials, time reversal symmetry alone is broken. The spatial inversion symmetry is related to   potentials that display relations such as $V(x)=   V(-x)$. The spatial inversion symmetry of systems  yields the constraints on both transmission and reflection amplitudes: $t^{(R)} (k) = t^{(L)} (k)$ and $r^{(R)}  (k)= r^{(L)} (k)$, which hold regardless if the potential is real or complex. The concept of reciprocity is distinct from time reversal symmetry, usually it refers to the equality in the signal received  when the source and detector are reversed, see Refs.~\cite{PhysRevA.78.064101,doi:10.1063/1.1704136,DILLON1968623}. In terms of potential operator, the reciprocity holds if   
\begin{equation}
\hat{V} = \hat{V}^T \label{ReciprocityVpot}
\end{equation}
 is satisfied. For a local potential $$\langle x' | \hat{V} | x \rangle = \delta (x-x') V(x)$$ regardless real or complex, Eq.(\ref{ReciprocityVpot}) is guaranteed. The  reciprocity symmetry leads to the constraint only on transmission amplitudes: $t^{(R)}(k) = t^{(L)} (k)$. In this subsection, we will   give a brief discussion on the reciprocity and spatial inversion symmetry for  local complex potentials. The discussion on combined  $\mathcal{P} \mathcal{T}$ symmetry will be  given separately in Sec.\ref{PTpotconstraints}.

\subsubsection{Reciprocity}
For a local complex potential, reciprocity symmetry is automatically satisfied: 
\begin{equation}
\langle x' | \hat{V} | x \rangle  = \langle x | \hat{V} | x' \rangle = \delta (x-x') V(x).
\end{equation}
One of the important consequences of reciprocity is that the transmission amplitudes for the right/left incident particles are identical,   see Refs.\cite{PhysRevA.78.064101,PhysRevA.64.042716,Mostafazadeh:2020iwo},
\begin{equation}
t^{(R)}(k) = t^{(L)} (k)     . \label{equaltRtL}
\end{equation}
Equality of right/left transmission amplitudes can be illustrated in a rather straightforward way. Using the Schr\"odinger equation Eq.(\ref{schrodingereq}), we obtain  
\begin{equation}
\frac{d}{d x}  W\left ( \Psi_k (x, k) ,  \Psi_k (x, -k)\right ) =0,
\end{equation}
where $W(f,g) = f g' - g f'$ refers to the Wronskian of two functions. Hence, we first conclude that the Wronskian of right/left propagating solutions of the Schr\"odinger equation, $\Psi_k (x, \pm k)$, doesn't depend on position $x$. Next using asymptotic behavior of wavefunctions in Eq.(\ref{asymwavabsorb}), we also find
\begin{equation}
 W  \left ( \Psi_k (x, k) ,\Psi_k (x, -k)   \right )  
 = 
 \begin{cases}
  - 2 i k t^{(R)} (k) ,  &     x \rightarrow + \infty ,   \\
   - 2 i k t^{(L)} (k) ,&    x \rightarrow - \infty .
  \end{cases}
\end{equation}
Together with  the fact that  Wronskian of right/left propagating solutions of the Schr\"odinger equation, $\Psi_k (x, \pm k)$, doesn't depend on position $x$ therefore yields the equality of right/left transmission amplitudes in Eq.(\ref{equaltRtL}).

\subsubsection{Spatial inversion}\label{Smatspatialinversion}
Next, let's consider a local   complex potential that display the spatial inversion symmetry,
\begin{equation}
V(x) = V(-x). \label{symmetricVpot}
\end{equation}
For an absorbing system, using LS Eq.(\ref{LSabsorb}) and symmetry of potential,
  we   find
   \begin{align}
 &  \Psi_{  k} (-x, -p)  \nonumber \\
 & = e^{     i p x}  - \frac{i m}{k}  \int_{-\infty}^\infty  d x'  e^{i k |x-x'|} V(x')  \Psi_{  k}  (-x', -p) ,  
 \end{align}
compared with Eq.(\ref{LSabsorb}),  hence we get
   \begin{align}
  \Psi_{  k} (-x, -p)  =  \Psi_{  k} (x, p).   \label{wavesymVpot}
 \end{align}
 Next, using the definition of scattering amplitude in Eq.(\ref{scatampabsorb})
  combined with symmetry relation of wave function given in Eq.(\ref{wavesymVpot}), we also find
  \begin{equation}
      f_k( -p'  ,   - p )  = f_k(  p' ,   p )  . \label{ampsymVpot}
 \end{equation}
Similar relations also hold for an emissive system as well.

Therefore,  the spatial inversion symmetric potential  yields
  \begin{equation}
  t^{(R)}(k)   = t^{(L)} (k) =t  (k),   \ \ \ \    r^{(R)}(k)    = r^{(L)} (k) = r  (k) ,
 \end{equation}
 and the $S$-matrix in parity basis becomes diagonal and requires only two independent complex functions, 
   \begin{equation}
  S^{(+/-)}(k)   =  \begin{bmatrix}  t  (k)   + r  (k)  &   0  \\   0 &  t  (k)   - r  (k)\end{bmatrix}       .
 \end{equation}
 Hence it is now possible to use two  real inelasticities and two real  phaseshifts  to parameterize  $S$-matrix,
    \begin{equation}
  S^{(+/-)}(k)   =  \begin{bmatrix}  \eta_+(k) e^{2 i \delta_+ (k) }  &   0  \\   0 & \eta_-(k) e^{2 i \delta_- (k) }   \end{bmatrix}       . \label{Smatparityparameterization}
 \end{equation}
 The unitarity relation  $$    \left [ S^{(+/-) }( -k) \right ]^T = \left [ S^{(+/-) }( k) \right ]^{-1}$$ adds  extra constraints for the elements of $S$-matrix defined below and above real $E$ axis,
    \begin{equation}
        \eta_\pm (-k) e^{2 i \delta_\pm (-k) }   =\eta^{-1}_\pm (k) e^{-2 i \delta_\pm (k) }    . \label{extraconstraints}
 \end{equation}

As $Im \hat{V} \rightarrow 0$,  dual systems become elastic and  $$\eta_\pm \stackrel{Im \hat{V} \rightarrow 0}{ \rightarrow} 1 ,$$
     extra constraints in Eq.(\ref{extraconstraints}) yield
    \begin{equation}
          \delta_\pm (-k)      \stackrel{Im \hat{V} \rightarrow 0}{ \rightarrow}   -  \delta_\pm (k)    ,
 \end{equation}
and
    \begin{align}
 & \widetilde{S}^{(+/-)\dag}  (k)  \stackrel{Im \hat{V} \rightarrow 0}{ \rightarrow}   S^{(+/-)\dag}  (k)  =   \begin{bmatrix}   e^{ - 2 i \delta_+ (k) }  &   0  \\   0 &  e^{ - 2 i \delta_- (k) }   \end{bmatrix}    .
 \end{align}
 The unitarity relation is  hence reduced to familiar form,
   \begin{equation}
S^{(+/-)\dag}  (k)   S^{(+/-) } (k)  = \mathbb{I}   .
 \end{equation}

\subsection{$\mathcal{P}\mathcal{T}$  symmetry}\label{PTpotconstraints}

For a local complex potential  that displays the combined $\mathcal{P}\mathcal{T}$ symmetry,
\begin{equation}
V^*(x) = V(-x),
\end{equation}
 the most intriguing part is that the  $\mathcal{P}\mathcal{T}$ symmetric potential imposes the symmetry constraints between  both an absorbing system and its dual system, which is different from the symmetry relations imposed by  symmetric potentials such as one in Eq.(\ref{symmetricVpot}). In the case of   symmetric potential in Eq.(\ref{symmetricVpot}), the symmetry constraints are only imposed on an absorbing and its dual system separately, see e.g. Eq.(\ref{wavesymVpot}) and Eq.(\ref{ampsymVpot}). In addition to $\mathcal{P}\mathcal{T}$ symmetry, since only local potential is considered in present work, the reciprocity symmetry is also satisfied automatically for dual systems, hence,  for an absorbing system, we find
\begin{equation}
f_k (k, k) = f_k (-k, -k) , \ \ \ \ t^{(R)} (k) = t^{(L)} (k) = t  (k).
\end{equation}
Similar relations hold for an emissive system as well.

\subsubsection{$\mathcal{P}\mathcal{T}$  symmetry constraints on wavefunctions and scattering amplitudes of dual systems}

Using LS equation Eq.(\ref{LSemissive}) combined with $\mathcal{P}\mathcal{T}$ symmetric potential, for an emissive system, we  thus  get
   \begin{align}
 & \widetilde{\Psi}_{  k} (-x,-p) \nonumber \\
 &= e^{     i  p x}    - \frac{i m}{ k}  \int_{-\infty}^\infty  d x'  e^{i k |x-x'|}  V(x') \widetilde{ \Psi}_{  k}  (-x',-p) ,  
 \end{align}
 compared with LS equation of absorbing system in Eq.(\ref{LSabsorb}),  we find 
  \begin{equation}
  \widetilde{\Psi}_{  k} (x, p)=  \Psi_{    k } (-x, - p), \ \ \ \ p = \pm k . \label{wavePTsymcond}
 \end{equation}
 Eq.(\ref{wavePTsymcond}) displays explicit symmetry relation between an absorbing system and its dual system imposed by  $\mathcal{P}\mathcal{T}$ symmetry.
Next, using definition of scattering amplitudes of dual systems in Eq.(\ref{scatampabsorb}) and Eq.(\ref{scatampregen}) combined with Eq.(\ref{wavePTsymcond}), we also find 
    \begin{equation}
  \widetilde{f}_k( p' ,   p )  =f_{k} ( -p' ,   -p ), \ \ \ \ (p',p) \in \pm k. \label{scattampPTsymcond}
 \end{equation}

Putting all together, for $\mathcal{P}\mathcal{T}$ symmetric dual systems, the plane wave basis wavefunctions  in dual systems are related by
  \begin{equation}
  \widetilde{\Psi}^*_{  k} (x, p)=   \Psi^*_{    k } (-x,  -p) =  \Psi_{    -k } (x, -p) , \label{PTandDualwavecond}
 \end{equation}
 and the scattering amplitudes are related by
    \begin{equation}
  \widetilde{f}^*_k( p' ,   p )  =f^*_{k} ( -p' ,   -p ) =- f_{-k} ( -p' ,   -p ).
 \end{equation}
In parity basis, the relations are given by
 \begin{equation}
  \widetilde{\Psi}^{(+/-) *}_{  k} (x)=  \pm \Psi^{(+/-) *}_{    k } (-x)  =     \Psi^{(+/-)}_{    -k } (x)   , 
 \end{equation}
  and
   \begin{equation}
  \widetilde{f}^{(+/-) *}_k( p'  ) = \pm f^{(+/-) *}_{k} ( -p'  ) =- f^{(+/-)}_{-k} ( -p'  ) ,  \ \ \ \ p'= \pm k .
 \end{equation}

\subsubsection{Parameterization of  $\mathcal{P}\mathcal{T}$ symmetric $S$-matrix} 

With the constraints of $\mathcal{R}$ symmetry for a local potential, now $S$-matrix for an absorbing system only depends on three complex functions: $t(k)$ and $r^{(R/L)} (k)$. The $\mathcal{P}\mathcal{T}$ symmetry put further constraints on its dual, using Eq.(\ref{scattampPTsymcond}), we find
 \begin{equation}
 \widetilde{S}^{(R/L) }  (k)  = \left [ S^{(R/L) } ( k) \right ]^T = \begin{bmatrix}  t  (k) &r^{(R) } (k) \\ r^{(L) } (k) & t  (k)  \end{bmatrix}   .
 \end{equation}
 Hence the unitarity relation for $\mathcal{P}\mathcal{T}$ symmetric dual systems now is given by
    \begin{equation}
 \widetilde{S}^{(R/L)\dag}  (k)   S^{(R/L) } ( k)=  S^{(R/L)* } ( k)   S^{(R/L) } ( k) = \mathbb{I}.
 \end{equation}

Next, let's illustrate the consequence of  $\mathcal{P}\mathcal{T}$ symmetry on $S$-matrix.  Working in  parity basis, $S$-matrix has the form of
\begin{equation}
 S^{(+/-)} (k)  
 =   \begin{bmatrix} 
A_+ (k)& B (k) \\
 - B(k) & A_- (k)
 \end{bmatrix}  ,   \label{PTSmatparity}
\end{equation}
where
\begin{align}
A_\pm (k)  &= t(k)  \pm \frac{r^{(R) }(k)  + r^{(L)} (k)}{2},  \nonumber \\
 B(k) & = \frac{r^{(R)  }(k) - r^{(L)} (k)}{2} .
\end{align}
The unitarity  relation  yields  three independent equations
\begin{align}
& |A_\pm (k)|^2 - |B (k)|^2  = 1,  \nonumber \\
& A_+ (k)^* B (k) + B^* (k) A_-(k)= 0 . \label{PTconstraint}
\end{align}
With three constraints in Eq.(\ref{PTconstraint}), the   independent functions in $S$-matrix   are now reduced to three real functions. Hence $\mathcal{P}\mathcal{T}$ symmetric  $S$-matrix can be parameterized by two   phaseshifts, $\delta_\pm (k)$,  and one  inelasticity, $\eta(k)$.  The solutions of Eq.(\ref{PTconstraint}) are  
\begin{align}
& A_\pm (k)= \eta  (k) e^{2 i \delta_\pm (k) }, \nonumber \\
& B(k)= i \sqrt{ \eta^2 (k)-1} e^{i \left (\delta^{(+)} (k) + \delta^{(-)} (k) \right )},   \ \ \ \ \eta (k) \geqslant 1. \label{PTSmatparam}
\end{align}
In terms of phaseshifts and inelasticity, the transmission and reflection amplitudes are given by
\begin{align}
t(k)  &= \eta  (k)  \frac{e^{2 i \delta_+ (k) } + e^{2 i \delta_- (k) }}{2}, \nonumber \\
r^{(R/L)} (k) &= \eta  (k)  \frac{e^{2 i \delta_+ (k) } - e^{2 i \delta_- (k) }}{2}  \nonumber \\
&\pm i \sqrt{ \eta^2 (k)-1} e^{i \left (\delta^{(+)} (k) + \delta^{(-)} (k) \right )} . \label{tandrampsPTsym}
\end{align}
 The unitarity relation  $$    \left [ S^{(R/L) }( -k) \right ]^T =   S^{(R/L)  *}( k) $$ adds  extra constraints for the elements of $S$-matrix defined below and above real $E$ axis,
    \begin{equation}
       t(-k) = t^*(k) , \ \ \ \ r^{(L)} (-k) = r^{(R)*} (k) . \label{extraconstraintsonPTsym}
 \end{equation}

The parameterization of $\mathcal{P}\mathcal{T}$ symmetric $S$-matrix in  Eq.(\ref{PTSmatparity}) and Eq.(\ref{PTSmatparam}) resembles the parameterization of $S$-matrix for a coupled-channel system of real potential scattering, see e.g. Refs.~\cite{Guo:2013vsa,Guo:2012hv}. However, in real potential scattering theory, the $S$-matrix of a two-channel  systems for a single partial wave, e.g.  $S$-wave, has the symmetric form of 
\begin{equation}
 S 
 =   \begin{bmatrix} 
A_1 & B   \\
  B  &  A_2  
 \end{bmatrix}  =  \left [ S   \right ]^T .     
\end{equation}
The constraints along diagonal direction of a symmetric $S$-matrix   become
\begin{equation}
 |A_{1/2}  |^2 + |B |^2  = 1.    \label{RealSmatconstraint}
\end{equation}
Hence in terms of    phaseshifts and   inelasticity, the $S$-matrix of two-channel system for a real potential scattering is parameterized by
\begin{equation}
 S     =   \begin{bmatrix} 
 \eta  e^{2 i \delta_1   } &  i   \sqrt{  1-\eta^2    }e^{ i \left ( \delta_1  + \delta_2   \right )} \\
 i   \sqrt{ 1- \eta^2  }e^{ i \left ( \delta_1   + \delta_2   \right )}  &  \eta   e^{2 i \delta_2   } 
 \end{bmatrix}  ,
\end{equation}
where the constraint equation in Eq.(\ref{RealSmatconstraint}) results that the value of inelasticity is in the range of $\eta \in [0,1]$. On the contrary, in  $\mathcal{P}\mathcal{T}$ symmetric systems, anti-symmetric form of $S$-matrix along off-diagonal direction ultimately leads to $\eta \geqslant 1$. Since spatial inversion alone is not a good symmetry in  $\mathcal{P}\mathcal{T}$ symmetric systems, the mixing effect between parity basis solutions also contribute. The inelasticity  in $\mathcal{P}\mathcal{T}$ symmetric systems  hence describes the transition between  parity basis solutions, which resembles to the inelasticity in two-coupled real potential scattering system that is used to describe the transition between two channels.

\section{Spectral representation of Green's function in complex potential scattering theory}\label{spectralGreensec}

\subsection{Spectral representation of Green's function}

The biorthogonality of eigenstates of dual systems in Eq.(\ref{wavdualnorm}) suggests that the spectral representation of Green's function for an absorbing system is defined by
\begin{equation}
\hat{G}(E)  = \sum_\epsilon \frac{ | \Psi_\epsilon \rangle \langle \widetilde{\Psi}_\epsilon |}{E- \epsilon},  \label{spectralG}
\end{equation}
and $\hat{G}(E) $ satisfies differential equation,
\begin{equation}
(E - \hat{H})\hat{G}(E)  =  \mathbb{I} .  \label{Gabsorbdiffeq}
\end{equation}
Similarly for an emissive system, we    have
\begin{equation}
\hat{\widetilde{G}}(E)   = \sum_\epsilon \frac{ |\widetilde{ \Psi}_{\epsilon} \rangle \langle \Psi_{\epsilon} |}{E- \epsilon}, 
\end{equation}
and $\hat{\widetilde{G}}(E) $   satisfies differential equation
\begin{equation}
(E - \hat{H}^\dag) \hat{\widetilde{G}}(E)  =  \mathbb{I} , \label{Gemissivediffeq}
\end{equation}
and  also Dyson equation  
\begin{equation} 
 \hat{\widetilde{G}} (E)= \hat{G}_0 (E)+ \hat{G}_0 (E)   \hat{V}^\dag  \hat{\widetilde{G}} (E). \label{Dysoneqemissive}
\end{equation}

In general, the Dyson equations for both absorbing and emissive systems in Eq.(\ref{Dysoneq}) and Eq.(\ref{Dysoneqemissive}) respectively are direction dependent and non-reciprocal: transpose of Green's function is not identical to the Green's function itself. However, for the local potentials, reciprocity symmetry is guaranteed: $$\hat{H}^T = \hat{H},$$ and using Eq.(\ref{Gabsorbdiffeq}) and Eq.(\ref{Gemissivediffeq}), we can easily show that Green's functions are indeed reciprocity symmetric:
\begin{equation}
\hat{G}(E)  =\hat{G}^T(E), \ \ \ \  \hat{\widetilde{G}}(E)  =\hat{\widetilde{G}}^T(E).
\end{equation}
Therefore taking into consideration of reciprocity symmetry, now  $\hat{G}(E) $  and $\hat{\widetilde{G}}(E) $ are      related by 
\begin{equation}
\hat{G}(E) = \hat{\widetilde{G}}^*(E^*) .
\end{equation}

From spectral representation of $\hat{G}(E)$ in Eq.(\ref{spectralG}), we find
\begin{equation}
\langle x |  \hat{G}(E) | x \rangle  = \sum_\epsilon \frac{ \langle x | \Psi_\epsilon \rangle \langle \widetilde{\Psi}_\epsilon | x \rangle }{E- \epsilon},   
\end{equation}
hence the absorptive part of Green's function is given by discontinuity of Green's function across branch cut in complex $E$-plane,
\begin{equation}
 - Disc_E \langle x |  \hat{G}(E) | x \rangle  =  \pi \sum_\epsilon   \delta(E - \epsilon)   \langle x | \Psi_\epsilon \rangle \langle \widetilde{\Psi}_\epsilon | x \rangle .  
\end{equation}
Therefore we can   conclude that  in complex potential scattering theory,  (1) the imaginary part of Green's function is not the same as absorptive part of Green's function, the absorptive part of Green's function in general could be a complex function. However, with constraints under $\mathcal{P} \mathcal{T}$ symmetry, the absorptive part of Green's function is indeed real, this will be demonstrated below in Sec.\ref{absorptivepartGPTsym}; (2) the absorptive part of Green's function can no longer interpreted as density of states in complex potential scattering theory.

\subsection{Absorptive part of Green's function under $\mathcal{P} \mathcal{T}$ symmetry}\label{absorptivepartGPTsym}

    The spectral representation of Green's function is   explicitly  given by
 \begin{equation}
 \langle x | \hat{G} (E)  | x' \rangle  = \int_{-\infty}^\infty \frac{d q}{2\pi}   \frac{ \sum_{p= \pm q}  \Psi_{  q}  (x, p) \widetilde{  \Psi}^*_{   q}  (x', p) }{  E-  \frac{ q^2 }{2m}},
\end{equation}
where the wavefunctions are eigen-solution of LS equations in Eq.(\ref{LSabsorb}) and Eq.(\ref{LSemissive}).
The discontinuity of diagonal elements of Green's function is thus  
 \begin{align}
 & Disc_E \langle x | \hat{G} (E)  | x \rangle   \nonumber \\
 &=-        \frac{m}{ 2 k}   \sum_{p= \pm k}  \left [ \Psi_{  k}  (x, p) \widetilde{  \Psi}^*_{   k}  (x, p) + \Psi_{ - k}  (x, p) \widetilde{  \Psi}^*_{   - k}  (x, p)  \right ].
\end{align}
 Using  $ \mathcal{P}\mathcal{T} $ symmetric relations on wavefunctions in Eq.(\ref{PTandDualwavecond}), we thus find
 \begin{equation}
   Disc_E   \langle x | \hat{G} (E)  | x \rangle   =   -    \frac{m}{ k}   \sum_{p= \pm k}   Re \left [ \Psi_{  k}  (x, p)  \Psi^*_{   k}  (-x, -p) \right ]. \label{DiscGPTsym}
\end{equation}
Therefore, under the constraints of $\mathcal{P} \mathcal{T}$ symmetry, the absorptive part of diagonal elements of Green's function is a real function, however the positivity is not guaranteed.

\bibliography{ALL-REF.bib}

\end{document}